\documentclass[10pt,3p]{elsarticle}
\journal{arXiv}

\usepackage{amssymb}
\usepackage[fleqn]{amsmath}
\usepackage{subcaption}
\usepackage{algpseudocode}
\usepackage{algorithm}
\usepackage{setspace}
\usepackage{hyperref}
\usepackage{cancel}
\usepackage{graphicx}
\usepackage{caption}
\usepackage{color}
\usepackage{hyperref}
\usepackage{siunitx}
\usepackage{url}

\newcommand{\B}{\mathcal{B}}

\newcommand{\V}{\mathcal{V}}
\newcommand{\Sol}{\mathcal{S}}
\newcommand{\Hilb}{\mathcal{H}}
\newcommand{\bs}[1]{\boldsymbol{#1}}
\newcommand{\pt}{\partial}

\newcommand{\ufsce}{\frac{\pt \bs{U}^n_e}{ \pt \bs{\beta}}} 
\newcommand{\xifsce}{\frac{\pt \bs{\xi}^n_e}{ \pt \bs{\beta}}} 
\newcommand{\ufspe}{\frac{\pt \bs{U}^{n-1}_e}{ \pt \bs{\beta}}} 
\newcommand{\xifspe}{\frac{\pt \bs{\xi}^{n-1}_e}{ \pt \bs{\beta}}}
\newcommand{\ufsc}{\frac{\pt \bs{U}^n}{ \pt \bs{\beta}}}

\newcommand{\bzeta}{\bar{\bs{\zeta}}^e}

\newcommand{\Ie}{\bar{I}^e} 
\newcommand{\bebar}{\bar{\bs{b}}^e}

\newcommand{\F}{\bs{F}}
\newcommand{\f}{\bs{f}}

\newcommand{\bbeta}{\bs{\beta}}
\newcommand{\bzero}{\bs{0}}

\newcommand{\epsR}{\varepsilon_{\bs{R}}}
\newcommand{\epsC}{\varepsilon_{\bs{C}}}

\newcommand{\obj}{\mathcal{J}}
\newcommand{\objgrad}{\frac{d\mathcal{J}}{d\bs{\beta}}}

\begin{document}

\begin{frontmatter}

\title{Calibration of Elastoplastic Constitutive Model Parameters from Full-field Data with Automatic Differentiation-based Sensitivities}
\author[sandia]{D. Thomas Seidl\corref{correspondence}}
\author[sandia]{Brian N. Granzow}
\address[sandia]{Sandia National Laboratories \\
P.O. Box 5800 Albuquerque, NM 87185-1321, United States}
\cortext[correspondence]{Corresponding author, dtseidl@sandia.gov}


\begin{abstract}
We present a framework for calibration of parameters in elastoplastic constitutive models that is based on the use of automatic differentiation (AD). The model calibration problem is posed as a partial differential equation-constrained optimization problem where a finite element (FE) model of the coupled equilibrium equation and constitutive model evolution equations serves as the constraint. The objective function quantifies the mismatch between the displacement predicted by the FE model and full-field digital image correlation data, and the optimization problem is solved using gradient-based optimization algorithms. Forward and adjoint sensitivities are used to compute the gradient at considerably less cost than its calculation from finite difference approximations. Through the use of AD, we need only to write the constraints in terms of AD objects, where all of the derivatives required for the forward and inverse problems are obtained by appropriately seeding and evaluating these quantities. We present three numerical examples that verify the correctness of the gradient, demonstrate the AD approach's parallel computation capabilities via application to a large-scale FE model, and highlight the formulation's ease of extensibility to other classes of constitutive models.

\end{abstract}

\end{frontmatter}

\section{Introduction}
\label{sec:introduction}

Many scenarios of interest in science and engineering involve large, permanent deformations (e.g.\ vehicle crashes and metal forming). In continuum mechanics the finite deformation theory of plasticity enjoys an increased range of applicability over infinitesimal strain formulations, but its mathematical theory as well as computational implementation through finite elements is more challenging. The field of computational plasticity has matured and computing power has increased to the point that finite deformation elastoplastic models are  used in industry, government, and academic labs, although their efficient implementation on high-performance computing (HPC) platforms remains an active area of research.

In order for such models to provide credible predictions, the parameters in the constitutive models must be determined through a calibration procedure which amounts to formulating and solving an inverse problem. In a typical calibration experiment, a sample is deformed in a controlled manner and the deformation and resultant loads are measured. These measurements are then used to inform calibration of an analytical or computational model of the material under study. Traditionally, specialized test specimens such as dogbones for metals described by elastoplastic constitutive models have been used because their geometry justifies the use of assumptions that simplify the mathematical modeling. Namely, prior to necking, the stress and strain are uniaxial and constant in the gauge section, and can be calculated analytically given the applied force and cross-section area (stress) and an extensometer and initial gauge length (strain).  While these assumptions facilitate the calibration procedure, they can also increase the number of mechanical tests required to produce an adequate amount of data for accurate calibration. For example, calibration of viscoplastic constitutive models using dogbone specimens requires the execution of multiple tests at constant and distinct strain rates and/or temperatures.

The maturity and relative ease of application of digital image correlation (DIC) provides opportunities to advance beyond traditional test specimens and their associated assumptions \cite{hild2006digital, sutton2009image, hild2012comparison}. In a typical test, hundreds of optical images of a patterned surface of the sample are acquired. The DIC algorithm then produces a full-field (i.e.\ densely distributed over the DIC region-of-interest) point cloud of displacement measurements, which can contain tens to hundreds of thousands of points depending on camera resolution and user-defined DIC settings. Furthermore, the typical amount of noise in present in one of these point measurements is on the order of $0.01 - 0.05$ pixels \cite{reu2015camera}. The conversion from image to physical units depends on the resolution of the camera and physical dimensions of the field-of-view, but for a well-controlled measurement in a laboratory setting the noise floor in physical units can readily be sub-micron.  In fact, the quality of the displacement measurements are so good that one of the primary uses of DIC is to provide experimental data for validation of computational finite element (FE) models, i.e.\ the DIC measurements are treated as a ground truth.

Digital volume correlation (DVC), the three-dimensional extension of DIC, is being actively developed although its use in constitutive model calibration is less mature \cite{bay1999digital,buljac2018digital}. It has the advantage of providing displacements through a sample volume, but in contrast to DIC, it is often difficult or impossible to create an artificial pattern in the sample interior, especially without modifying the base material properties. Therefore, in most cases the microstructure itself serves as the pattern, which is typically not optimal for DVC algorithms \cite{bay2008methods,croom2019effect}. Further, the method of producing images must be capable of penetrating a volume, so imaging systems based on X-rays are commonly used. We do not consider calibration from DVC data in this paper, but extending our computational formulation to volumetric displacement data would be trivial.

As we will discuss shortly, the experimental mechanics community that developed DIC has led the way in using DIC data to calibrate constitutive models, and the most common means of doing so is through an optimization formulation that minimizes the difference between experimental measurements (e.g.\ DIC data and load) and FE model predictions. It is challenging, however, to fully capitalize on the rich deformation data provided by DIC, as the FE model has to have sufficient fidelity to accurately approximate the motion of the specimen (i.e.\ to control discretization error in the model) and represent the DIC data in regions of rapid variation. The optimization approaches favored in the literature (genetic algorithms and gradient-based methods that rely on finite difference approximations) require many solutions of the FE model (easily hundreds to thousands), which places a practical limit on the cost and ultimately fidelity of the model, to the point where the high-resolution DIC data is often down-sampled to a considerably coarser FE mesh.

In this paper we present a proof of concept for calibration of parameters in large-scale, finite deformation elastoplastic constitutive models from full-field deformation data using a deterministic partial differential equation (PDE)-constrained optimization formulation where the gradient of the objective function is obtained through forward and adjoint sensitivity analyses. Following \cite{michaleris1994tangent,kulkarni2007newton,alberdi2018unified}, we write the constraint equations as coupled \emph{residuals},  i.e.\ discretized weak forms for the \emph{global} equilibrium PDE and \emph{local} constitutive model evolution equations. A defining feature of our approach is the use of automatic differentiation (AD) to facilitate the computation of the many derivatives present in the formulation.

Automatic differentiation is a technique to computationally evaluate analytic derivatives (to machine precision) of a given function. In the context of PDE-constrained optimization, AD provides the ability to compute various required sensitivities based on implementations of the residuals (and objective function when its derivatives cannot be easily determined), {\textit{without requiring manual derivative calculations}}. These sensitivities include the derivatives of these quantities with respect to state variables (e.g.\ displacement, equivalent plastic strain) and constitutive model parameters. In particular, our approach utilizes forward mode AD to obtain derivative quantities at the \emph{element level}. Utilizing AD in this manner facilitates the parallelization of required computations \cite{pawlowski2012automating1, pawlowski2012automating2}. For more details on AD, we refer readers to the canonical book \cite{griewank2008evaluating}. Our computational implementation uses the \texttt{C++} operator-overloading AD software package \texttt{Sacado} \cite{phipps2012efficient}.

After presenting the formulation of the inverse problem, we provide three numerical calibration examples with synthetic DIC data and note that several algorithmic refinements would be required to make our framework suitable for calibration with experimental full-field data. The framework described in this paper, however, is amenable to such extensions. We also remark that our use of AD makes the implementation of other types of constitutive models (e.g.\ viscoplastic, kinematic hardening, damage) easier than in traditional approaches. Additionally, the model calibration problem shares many similarities to optimal design problems studied in the computational mechanics literature such that this work may be useful in those fields as well.

In the remainder of the introduction, we review existing approaches to elastoplastic model calibration from full-field data and discuss how our work relates to studies in the PDE-constrained optimization and computational plasticity literature that address finite deformation constitutive models.

\subsection{Constitutive Model Calibration for Elastoplastic Materials from Full-field Data}

The two primary approaches in the experimental mechanics community for calibration from full-field DIC data are known as finite element model updating (FEMU) \cite{avril2008overview} and the virtual fields method (VFM) \cite{grediac2006virtual, pierron2012virtual}. In the VFM, the objective function is a squared mismatch between internal and external virtual work or power instead of measured and predicted displacement or strain subject to PDE constraints. We will not discuss VFM further, save to say it has been used to calibrate parameters in finite deformation elastoplastic constitutive models \cite{rossi2012identification,kramer2014implementation,rossi2016application,Jones2018,jones2019investigation,martins2019calibration,marek2020experimental}.

In FEMU, an objective function that quantifies the weighted mismatch between model predictions and corresponding experimentally-measured quantities of interest is minimized by iteratively updating the parameters of a FE model using an optimization algorithm. A basic objective function contains data from load-displacement curves, but more sophisticated objective functions employ full-field displacements or strains obtained from DIC. The formulation presented in this paper could be viewed as a FEMU method, the defining difference being that in the FEMU literature finite differences are used to compute the gradient of the objective function or derivatives of the displacement field with respect to the model parameters. This approach to differentiation allows the FE model to be treated in a black-box fashion and thus encourages the use of off-the-shelf software. We avoid the use of finite difference approximations in our work and instead employ forward and adjoint sensitivity analyses. While these techniques require intrusive, non-standard modifications of the FE solver, they are significantly more computationally efficient, as we will demonstrate in this work.

Integrated DIC \cite{rethore2013robust, mathieu2015estimation, neggers2017improving, ruybalid2016comparison, neggers2017reducing, bertin2016integrated, neggers2018big, neggers2019simultaneous} and DVC \cite{hild2016toward, buljac2018calibration} approaches that combine the displacement and constitutive model parameter estimation problems have been thoroughly investigated in the literature. In these formulations the objective function is the same one used in standard DIC, a least-squares formulation based on the \textit{optical residual} that quantifies the mismatch between a reference and deformed image warped by the DIC displacement estimates. The FE model is introduced as a constraint, and most approaches use a Gauss-Newton algorithm to solve the minimization problem. This method requires the computation of displacement sensitivity matrices, and these are also typically obtained using finite differences. Integrated DIC often makes use of global DIC methods, where a FE basis is used to represent the displacement field, and it is often the case that the same basis is used for the FE model and DIC displacements, although this isn't strictly necessary.

A notable drawback of the reliance on finite difference approximations is the introduction of scaling with dimension of the parameter space. A full nonlinear FE solve is needed to compute each component of the gradient through a forward finite difference approximation. This limitation restricts the practical size of the parameter space, such that most studies consider homogeneous materials described by a single constitutive model with 5-10 parameters. This scaling can also reduce the ability of the FE model to take full advantage of the high temporal and spatial fidelity of the DIC data when cost of solving the FE model on a ``fine-enough'' mesh is prohibitive.

Finally, due to the expense of the finite deformation elastoplastic FE models and high-dimensional nature of DIC data (spatial and temporal), Bayesian approaches to parameter calibration, where probability distributions for the parameters are obtained, are to date relatively rare. Deterministic calibration, where point estimates of the most likely values of the parameters are calculated given the experimental data, is more common. We do not consider Bayesian inverse formulations in this work, but the computationally-efficient gradient calculation methods we describe could be useful for the construction of surrogate models for emulator-based Bayesian inference and Markov-chain Monte Carlo algorithms that can take advantage of this information.

\subsection{PDE-constrained Optimization in Finite Deformation Plasticity}

We now turn to the studies of PDE-constrained optimization in the computational mechanics literature where forward and adjoint sensitivity analyses have been applied to finite deformation elastoplastic constitutive models for the purposes of model calibration and optimal design. Each of these articles invariably contains pages of analytic expressions for the derivatives (or related intermediate quantities) that we obtain through the application of AD to element-level residuals.

Parameter identification for finite deformation elastoplastic constitutive models using a principal space formulation has been presented in \cite{mahnken1997parameter, mahnken1999aspects, mahnken2000comprehensive, mahnken2002theoretical}. The gradient is analytically derived for the continuous form of the governing equations and then discretized (an \textit{optimize-then-discretize}  approach).  The numerical results in these studies focus on tensile specimens with measurements of displacement and load, and the objective function quantities the mismatch of point measurements of displacement that are either randomly chosen over a subset of the mesh or located in the necking region. 

Numerical formulations for finite deformation plasticity with applications to metal-forming \cite{badrinarayanan1996sensitivity,srikanth2000shape,zabaras2000continuum, srikanth2001updated, ganapathysubramanian2002continuum, ganapathysubramanian2003computational, zabaras2003continuum, acharjee2006continuum}  and optimal design of multiscale materials \cite{acharjee2003proper,ganapathysubramanian2004design, ganapathysubramanian2004deformation, sundararaghavan2005synergy, sundararaghavan2006design, sundararaghavan2008multi}  have also been developed using an \textit{optimize-then-discretize} approach based on forward sensitivity analysis. Notably, many of these formulations incorporate additional constraints such as contact or thermal physics and analytically derive the coupled linear systems used to compute the sensitivities needed for gradient evaluation.

In topology optimization the goal is to produce a structural design that satisfies an optimality criterion \cite{sigmund2013topology, liu2018current} given a volume constraint. Earlier studies focused on linear elastic models and minimum compliance formulations, but today authors consider elastoplastic constitutive models and other objective functions such as those that quantify energy dissipation. In topology optimization the design variable is typically a density field that has been ``filtered'' to produce $0$ or $1$ values (i.e.\ present or not present) within a given element.

Various authors have considered topology optimization problems for small-strain plasticity \cite{nakshatrala2015topology, wang2017efficient, zhang2017topology, li2017energy, li2017topologydamage, amir2017stress, alberdi2017topology, li2017fracture, alberdi2018unified, li2018failure, alberdi2019design},  but only a few have addressed finite deformation plasticity \cite{wallin2016topology, ivarsson2018topology, alberdi2019bi, alberdi2020optimized, ivarsson2020topology}. We also note that 3D numerical examples are not present in the latter, presumably due to the computational expense and complexity of implementation of these formulations. 

Like model calibration, topology optimization is formulated as a PDE-constrained optimization problem. The governing equations can be the same as in our approach, but the objective function is not. The optimization algorithm known as the method of moving asymptotes (MMA) \cite{svanberg1987method} is often used to solve these problems, and a detailed description of this algorithm is not relevant to our work. What is important to mention is that the MMA requires the derivative of the objective function with respect to the design parameters, and in this paper we provide an AD-based recipe for computing this quantity using forward or adjoint sensitivities. In topology optimization most authors use an adjoint-based approach to gradient calculation because the design variable is a heterogeneous material density and therefore is directly related to mesh size such that the dimension of the parameter space makes adjoint approaches the only viable option.

More specifically, most authors follow the adjoint-based approach described in \cite{michaleris1994tangent} to compute the gradient of the objective function, which was originally presented in the general context of transient coupled nonlinear systems.  More recently, \cite{alberdi2018unified} presented a unification of this framework for topology optimization of path-dependent materials that shares both the global-local split of the governing equations and \textit{discretize-then-optimize} strategy followed in our work.

While AD makes a few appearances in the topology optimization literature, to date it has seen more application in fluid mechanics \cite{rokicki2016adjoint, mishra2018gpu, dilgen2018density, dilgen2018topology}.  Its use in solid mechanics
is limited to 1D wave propagation \cite{norgaard2017applications} and more recently fluid-structure interaction \cite{gomes2020aerodynamic}.

Lastly, the work of \cite{korelc2009automation, korelc2014closed, korelc2016automation} is notable in that it describes fully-automated approach that utilizes source transformation AD for forward sensitivity analyses in computational plasticity, but it does not consider its use in constitutive model calibration problems. In these approaches AD is used to compute both the residual vector and its associated tangent matrix from a single scalar function that is derived from a strain-energy density function and the equations that govern plastic deformation, while in our formulation AD implementations of coupled residuals are the fundamental objects from which tangent matrices are obtained.

\subsection{AD for Computational Plasticity}

The use of AD in the forward problem (i.e.\ find displacements given model parameters) for finite deformation plasticity has been explored \cite{korelc2009automation, chen2014automatic, rothe2015automatic, li2017simulation}. Sometimes authors will compare the cost to non-AD based approaches, but it is not critical to belabor this comparison because the computational efficiencies of AD implementations vary. A primary benefit of AD in this context is that it makes the use of more complex mechanical models easier by automating the computation of tangent matrices needed for nonlinear solves.

We formulate the forward problem as a set of two coupled nonlinear residuals: one that is a global object (i.e.\ over all nodes) and the other local in the sense that it is restricted to a specific integration point (or element when there is a single integration point in the element as occurs in the models in this work) in the spirit of the topology optimization framework for coupled problems given in \cite{alberdi2018unified}. Thus our solution of the forward problem is based upon an AD-based version of the general frameworks presented in \cite{michaleris1994tangent,kulkarni2007newton} that utilize Newton solves for each set of nonlinear equations which ensures the consistency of the global and local tangent stiffness matrices. 

The manual derivation of these tangent matrices for complex constitutive models can be quite challenging, and we use AD to automate this task.  In our framework all that is required to solve full-field constitutive model calibration problems is an implementation of the global and local residuals and objective function (possibly without AD if it a ``nice'' function like a quadratic), which makes its extension to other classes of constitutive models relatively straightforward.

\subsection{Contributions}

Our work provides contributions in the following four areas:

\begin{itemize}

\item The FEMU approach to model calibration of finite deformation elastoplastic constitutive models from full field data favored by the experimental mechanics community relies on finite difference approximations, and we use forward and adjoint sensitivity analyses to obtain the gradient with considerably less computational cost.

\item While forward and adjoint sensitivity analyses have been performed for PDE-constrained optimization problems that involve finite deformation elastoplastic constitutive models with isotropic yield functions in the computational mechanics literature, this work is the first to do so through the application of AD to element-level residuals for the purposes of model calibration. The adjoint version of our formulation is similar to that taken by the topology optimization community, and with minor alterations our approach could be applied to such problems.

\item The HPC aspects of our approach, although not fully developed in this paper, could increase the utilization of high-fidelity DIC measurements beyond the amount used in the literature today by increasing the spatial and temporal resolution of the FE model to a level that is commensurate with the full-field data.  Similarly, if our approach is adapted for design optimization, it could enable large-scale optimization of 3D elastoplastic structures.

\item
Calibration of parameters in anisotropic yield models from full-field deformation data via finite difference-based FEMU and
integrated DIC techniques has been demonstrated for small-strain plasticity \cite{lecompte2009parameter, teaca2010identification, bertin2017identifiability} and finite membrane strain shell models \cite{wang2016anisotropic, coppieters2018inverse}. Topology
optimization of small-strain elastoplastic structures with anisotropic yield models has been investigated in \cite{zhang2017topology, mirzendehdel2018strength, alberdi2019design} using adjoint-based methods. We believe that this work contains the first application
of forward and adjoint sensitivities to a 3D finite deformation elastoplastic FE model with an anisotropic yield function.

\end{itemize}

\subsection{Outline}
The remainder of the paper is organized as follows. First, in section \ref{sec:forward} we describe the governing equations for a finite deformation elastoplastic constitutive model and their finite element discretization. Next, in section \ref{sec:inverse} we present an optimization-based formulation of the inverse problem and three methods for obtaining the gradient of the objective function. Results for three example problems that utilize synthetic DIC data are contained in section \ref{sec:results}. Section \ref{sec:discussion} contains a discussion of some of the limitations of our approach that would need to be addressed when using experimental DIC data for calibration. Conclusions are given in section \ref{sec:conclusion}.

\section{Finite deformation Forward Model}
\label{sec:forward}

In this section we describe the forward model. It naturally breaks into two pieces: the equilibrium PDE and the constitutive equation evolution equations. Upon discretization the former equation is satisfied in \emph{global} sense while latter are a system of \emph{local} equations that hold at each numerical integration or Gauss point.

\subsection{Equilibrium equation}

The balance of linear momentum in the absence of inertial terms and body
forces can be written as

\begin{gather}
\begin{cases}
\begin{aligned}
- \nabla \cdot \bs{P} &= \bs{0}, &&\text{in} \;\; \B, \\
\bs{u} &= \bs{g}, &&\text{on} \;\; \Gamma_g, \\
\bs{P} \cdot \bs{N} &= \bs{h}, &&\text{on} \;\; \Gamma_h.
\end{aligned}
\end{cases}
\label{eq:lin_momentum}
\end{gather}

\noindent Here, $\B \in \mathbb{R}^{3}$ denotes a three-dimensional open bounded domain. The boundary $\Gamma$
of $\B$ is assumed to be sufficiently smooth and can be decomposed such that
$\Gamma = \Gamma_g \cup \Gamma_h$ and $\Gamma_g \cap \Gamma_h = \varnothing$. The displacement field $\bs{u}$ is expressed as function of the coordinates in the reference configuration $\bs{X}$.
Additionally,  $\bs{P} := J \bs{\sigma} \bs{F}^{-T}$ denotes the
first Piola-Kirchhoff stress tensor, where $\bs{F} := \frac{\pt \bs{u}}{\pt \bs{X}} + \bs{I}$ denotes the deformation
gradient, $\bs{I}$ denotes second-order identity tensor, $J := \text{det}\left[\bs{F}\right]$ denotes the determinant of the deformation
gradient, and $\bs{\sigma}$ denotes the Cauchy stress tensor. Finally, $\bs{N}$ denotes
the unit outward normal to the surface $\Gamma$ in the reference configuration, $\bs{g}$ denotes an applied
external displacement, and $\bs{h}$ denotes an applied traction.

\subsection{Constitutive Equations}

We consider the finite deformation elastoplastic constitutive model outlined in Chapter 9 of Simo and Hughes
\cite{simo2006computational} with some modifications that will be discussed shortly.
 For the sake of brevity, we perform a high-level
review of this constitutive model and refer readers to \cite{simo2006computational,borden2016phase}
for further details. For the purposes of the current discussion, we make two
important notes about this constitutive model. First, the model
assumes an \emph{intermediate stress-free configuration}, which
leads to a multiplicative decomposition $\bs{F} = \bs{F}^e \bs{F}^p$ of the deformation gradient into
elastic and plastic components, $\bs{F}^e$ and $\bs{F}^p$, respectively.
Second, we make the assumption that the plastic deformation
is volume preserving, as is common in $J_2$ flow theory, such that
$J = J^e := \text{det}\left[\bs{F}^e\right]$ and
$J^p := \text{det}\left[\bs{F}^p\right] = 1$.

\vspace{1em}

\noindent \emph{Elastic Response:} Below the yield strength,
the material response is assumed to be
hyperelastic, such that

\begin{gather}
\begin{aligned}
\bs{\tau} &= \bs{s} - J p \bs{I} ,  \\
-p &:= \frac{\kappa}{2} (J^2 - 1) / J ,\\
\bs{s} &:= \text{dev} \left[ \bs{\tau} \right] = \mu \ \text{dev} \left[ \bar{\bs{b}}^e \right] = \mu \bzeta.
\label{eq:hyperelastic}
\end{aligned}
\end{gather}

\noindent Here, $\kappa = \frac{E}{3(1 - 2 \nu)}$ is the \emph{bulk modulus},
$\mu = \frac{E}{2(1 + \nu)}$ is the \emph{shear modulus}, $E$ denotes \emph{Young's modulus}, $\nu$ denotes \emph{Poisson's ratio},
$\bar{\bs{b}}^e := J^{-\frac{2}{3}} \bs{b}^e$ is the volume-preserving part of the elastic left Cauchy-Green strain tensor $\bs{b}^e := \bs{F}^e \left({\bs{F}^e}\right)^T$,
$\bs{\tau} := J \bs{\sigma}$ is the Kirchhoff stress tensor,
$p$ is the pressure, and $\bs{s}$ is the deviatoric part of the Kirchhoff stress tensor, where
$\text{dev} \left[ \cdot \right] := \left[\cdot\right] - \frac13 \text{tr} \left[\cdot \right] \bs{I}$.
The deviatoric and spherical parts of $\bar{\bs{b}}^e$ are represented by $\bzeta$ and $\Ie$, respectively, such that
$\bar{\bs{b}}^e = \bzeta + \Ie \bs{I}$.

\vspace{1em}

\noindent \emph{Yield Condition:}
A Mises-Huber yield condition with isotropic linear and/or Voce-type hardening results in a yield function of the form:

\begin{gather}
\begin{aligned}
f &:=  \| \bs{s} \| - \sqrt{\frac{2}{3}} \sigma_y, \\
\sigma_y &= Y + K \alpha + S \left( 1 - \exp \left(-D \alpha \right) \right).
\end{aligned}
\label{eq:yield_surface}
\end{gather}

\noindent Here $Y$ denotes the material's \emph{yield strength},
$K$ denotes a \emph{linear hardening modulus},
$S$ denotes a \emph{hardening saturation modulus},
$D$ denotes a \emph{hardening saturation rate},
and $\| \cdot \|$ applied to a second-order tensor denotes the
Frobenius norm. The equivalent plastic strain $\alpha$ is a non-negative internal variable that measures the
amount of plastic deformation experienced by the material at a point. The yield function satisfies
$f \leq 0$, where $f < 0$ implies an elastic material response and
$f = 0$ implies a plastic material response.  

Our parameterization of the constitutive model utilizes six independent material parameters: $E$, $\nu$, $K$, $Y$, $S$ and $D$. These constants, $\bs{\beta} = [E, \nu, Y, K, S, D]$, are the optimization parameters in constitutive model calibration.

\vspace{1em}

\noindent \emph{Isotropic Hardening and Loading/Unloading Conditions:}
We assume that the evolution of the equivalent plastic strain
$\alpha$ is governed by a hardening law of the form

\begin{gather}
\dot{\alpha} = \sqrt{\frac23} \dot{\gamma}.
\label{eq:eqps_evolution}
\end{gather}

\noindent Further, we assume that the plastic multiplier $\dot{\gamma}$ is
subject to the Kuhn-Tucker loading/unloading conditions:

\begin{gather}
\begin{aligned}
\dot{\gamma} &\geq 0, \\
f &\leq 0, \\
\dot{\gamma} f &= 0.
\end{aligned}
\end{gather}

\noindent \emph{Associative Flow Rule:} We consider an associative flow
rule obtained by applying the principle of maximum plastic dissipation:

\begin{gather}
\begin{aligned}
\text{dev} \left[ L_v \left( \bs{b}^e \right) \right] &=
\text{dev} \left[ \bs{F} \dot{\bs{G}}^p \bs{F}^T \right] = 
- \frac23 \dot{\gamma} \, \text{tr} \left[ \bs{b}^e \right] \bs{n}, \\
\bs{n} &:= \frac{\bs{s}}{\| \bs{s} \|}.
\label{eq:flow_rule}
\end{aligned}
\end{gather}

\noindent where $L_v \left( \bs{b}^e \right)$ is the Lie derivative of elastic left Cauchy-Green strain tensor, $\bs{G}^p := ({\bs{F}^p}^T \bs{F}^p)^{-1}$ is the
inverse of the plastic right Cauchy-Green strain tensor, and $\bs{n}$ is the yield surface normal. As noted in \cite{borden2016phase}, this flow rule only specifies $\text{dev} \left[ L_v \left( \bs{b}^e \right) \right]$. We 
must impose the isochoric plastic deformation assumption

\begin{gather}
\text{det} \left[ \bs{b}^e \right] = (J^e)^2 =  J^2, \label{eq:isochoric}
\end{gather}

\noindent to determine the spherical part of $\bs{b}^e$. 

\vspace{1em}

\subsection{Stabilized Finite Element Formulation}

Let the domain $\B$ be decomposed into $n_{el}$ non-overlapping
finite element subdomains $\B_e$ such that
$\B = \cup_{e=1}^{n_{el}} B_e$ and $B_i \cap \B_j = \varnothing$
if $i \neq j$. For our purposes, we consider tetrahedral elements. Let $\V_u$, $\V_w$, and $\V_p$
denote the finite dimensional function spaces defined as:

\begin{gather}
\begin{aligned}
\V_u &:= \{ \bs{u} : \bs{u} \in \Hilb^1(\B)^3, \;
\bs{u} = \bs{g} \; \text{on} \; \Gamma_g, \;
\bs{u} \bigr|_{\B_e} \in \mathbb{P}^1(\B_e)^3 \}, \\
\V_w &:= \{ \bs{w} : \bs{w} \in \Hilb^1(\B)^3, \;
\bs{w} = \bs{0} \; \text{on} \; \Gamma_g, \;
\bs{w} \bigr|_{\B_e} \in \mathbb{P}^1(\B_e)^3 \}, \\
\V_p &: =\{ p : p \in L^2(\B), \;
p \bigr|_{\B_e} \in \mathbb{P}^1(\B_e) \}. \label{eq:fe_spaces}
\end{aligned}
\end{gather}

\noindent Here $\Hilb^1$ denotes the space of square-integrable functions
with square-integrable first derivatives, $L^2$ denotes the
space of square-integrable functions, and $\mathbb{P}^1(\B_e)$
denotes the space of piecewise linear polynomials over elements
$\B_e$, $e = 1, 2, \dots, n_{el}$.

Following the approach of Maniatty et al.
\cite{klaas1999stabilized, maniatty2002higher, ramesh2005stabilized},
we utilize a stabilized Petrov-Galerkin finite element method
obtained by multiplying the equilibrium equation \eqref{eq:lin_momentum}
by a perturbed weighting function of the form
$\bs{w} + \tau_e \bs{F}^{-T} \nabla q$, multiplying the pressure
equation in \eqref{eq:hyperelastic} by a weighting function
$q$, and integrating over the domain $\B$. Here
$\bs{w} \in \V_u$ is a displacement weighting function,
$q \in \V_p$ is a pressure weighting function,
and $\tau_e = \frac{h_e^2}{2 \mu}$ is a mesh-dependent stabilization
parameter, where $h_e := \text{meas}(\B_e)$ is a characteristic size
of a mesh element.

Let $\Sol := \V_u \times \V_p$, $\V := \V_w \times \V_p$,
$\bs{U} := [\bs{u}, p]$, and $\bs{W} := [\bs{w}, q]$. We refer to $\bs{U}$ as the \emph{global state variables}.
Performing integration by parts on \eqref{eq:lin_momentum}
and using the relations in \eqref{eq:hyperelastic}, it can be shown
\cite{klaas1999stabilized, maniatty2002higher, ramesh2005stabilized,
granzow2018adjoint} that this strategy leads to the stabilized
finite element formulation: find $\bs{U} \in \Sol$ such that for
all $\bs{W} \in \V$

\begin{gather}
\begin{split}
&\int_{\B} (\mu \bzeta \bs{F}^{-T}) : \nabla \bs{w} \, \text{d} V -
\int_{\B} (J p \bs{F}^{-T}) : \nabla \bs{w} \, \text{dV} -
\int_{\B} \left[ \frac{p}{\kappa} + \frac{1}{2J}(J^2 - 1) \right] q \, \text{d} V - \\
&\sum_{e=1}^{n_{el}} \int_{\B_e} \tau_e (J \bs{F}^{-1} \bs{F}^{-T}) : (\nabla p \otimes \nabla q) \, \text{d} V -
\int_{\Gamma_h} \bs{h} \cdot \bs{w} \,\, \text{d} A = 0.
\end{split} \label{eq:eq_res}
\end{gather}

\noindent We note our use of linear shape functions allows for single-point Gauss quadrature for numerical integration of each term in \eqref{eq:eq_res} except for a single quadratic term $\frac{pq}{\kappa}$ that requires a four-point integration rule.

We use $\bs{R}$ to denote the semi-discrete nonlinear system of algebraic
equations (i.e.\ the \emph{global residual}) that arises from \eqref{eq:eq_res}. In the next subsection we will define the temporal discretization and describe $\bs{C}$, the fully-discrete equations for the evolution of the constitutive model (i.e.\ the \emph{local residual}).

\subsection{Discrete Constitutive Equation Evolution}

The constitutive model expressed by equations \eqref{eq:yield_surface}-\eqref{eq:isochoric} describes the time evolution of the local state variable $\alpha$ and the kinematic variable $\bs{b}^e$. We collect these variables into a vector $\bs{\xi} := \left\lbrace \bzeta, \Ie, \alpha \right\rbrace$ that represents the \emph{local state variables} and note that we have transformed $\bs{b}^e$  into $\bar{\bs{b}}^e$ and separated it into its spherical and deviatoric components. Only one of these variables, $\bzeta$, appears in \eqref{eq:eq_res}, and it suffices to treat its spatial discretization as being constant within a given element. Therefore, we use a piecewise-constant discretization for each locate state variable, which can be integrated exactly using a single Gauss point in each element.

For temporal discretization we utilize a backward Euler scheme in pseudo-time according to load step increments $[t^{n-1}, t^{n}]$ for a series of $n = 1, \ldots, N_L$ load steps with uniform unit spacing so that that $\Delta t = 1$. The backward Euler approximation of the time derivative of a function $u$ at load step $n$ is 

\begin{gather}
\frac{\partial u(\bs{x}, t^n)}{\partial t} \approx u(\bs{x}, t^n) - u(\bs{x}, t^{n-1}) = u^n - u^{n-1}. 
\end{gather}

At a given load step $n$ we first compute a trial value of the deviatoric part of the Kirchhoff stress $\bs{s}^n_{\text{trial}}$ and use it to evaluate the yield function to see if plastic flow has occurred (i.e.\ $f^n_{\text{trial}} > 0$). This result then determines the form of the nonlinear system of algebraic equations for the fully-discrete constitutive equation evolution residual $\bs{C}^n$.

Before defining the trial variables we introduce the relative deformation gradient $\bs{f}^n$ and its volume-preserving form $\bar{\bs{f}}^n$ and restate the deviatoric-spherical decomposition of $(\bebar)^n$:

\begin{gather}
\begin{aligned}
\f^n &:= \F^n  \left(\F^{n-1} \right)^{-1}, \\
\bar{\f}^n &:= \text{det} \left( \f^n \right)^{-\frac{1}{3}} \f^n, \\
\left(\bebar\right)^n &= \left(\bzeta\right)^n + \left( \Ie \right)^n \bs{I} . \label{eq:disc_kin_internal}
\end{aligned}
\end{gather}

\noindent The variables for the trial step are

\begin{gather}
\begin{aligned}
\alpha^n_\text{trial} &= \alpha^{n-1}, \\
\left(\bebar \right)^n_{\text{trial}} &= \bar{\f}^n \left( \left( \bzeta \right)^{n-1} + (\Ie)^{n-1} \bs{I} \right) \left( \bar{\f}^n \right)^T, \\
\left(\bzeta \right)^n_{\text{trial}} &= \text{dev} \left[ \left(\bebar\right)^n_{\text{trial}}  \right], \\
\left(\Ie \right)^n_{\text{trial}} &= \frac{1}{3} \text{tr} \left[ \left(\bebar\right)^n_{\text{trial}}  \right], \\
\bs{s}^n_{\text{trial}}  &= \mu \left(\bzeta \right)^n_{\text{trial}}.
\end{aligned} 
\end{gather}

If $f^n_\text{trial} \leq 0$ then we set the local state variables at the current load step to their trial values. In this case the form of $\bs{C}^n$ is particularly simple:

\begin{gather}
\begin{cases}
\begin{aligned}
\left( \bzeta \right)^{n} - \left(\bzeta \right)^n_{\text{trial}}   &= \bs{0}, \\[8pt]
\left( \Ie \right)^{n} - \left(\Ie \right)^n_{\text{trial}}   &= 0, \\[8pt]
\alpha^n - \alpha^{n}_\text{trial}  &= 0.
\end{aligned}
\end{cases} \label{eq:Cn_elastic}
\end{gather}

The form of $\bs{C}^n$ when $f^n_{\text{trial}} > 0$ is given by the discrete versions of the equivalent plastic strain evolution equation \eqref{eq:eqps_evolution},  flow rule \eqref{eq:flow_rule}, isochoric plastic deformation constraint \eqref{eq:isochoric}, and yield function \eqref{eq:yield_surface}. Starting with the first of these, the discrete counterpart of the plastic multiplier $\dot{\gamma}$ is traditionally denoted by $\Delta \gamma$. Backward Euler discretization of \eqref{eq:eqps_evolution} yields

\begin{gather}
\alpha^n - \alpha^{n-1} = \sqrt{\frac23}  \Delta \gamma. \label{eq:disc_eqps_evolution}
\end{gather}

It is more convenient to work with a time-discretized version of \eqref{eq:flow_rule} and \eqref{eq:isochoric}  that has been multiplied on both sides by $\left(J^n\right)^{-\frac23}$:

\begin{gather}
\begin{aligned}
\text{dev} \left[ \bar{\bs{F}}^n \left\lbrace \left( \bs{G}^p \right)^{n} - \left( \bs{G}^p \right)^{n-1} \right\rbrace \left( \bar{\bs{F}}^n \right)^T \right] &= - \frac23 \Delta \gamma 
\text{tr} \left[ \left( \bebar \right)^n \right] \bs{n}^n,  \\
\text{det} \left[ \left(\bebar \right)^n \right] &= 1.  \label{eq:volume_pres_forms}
\end{aligned}
\end{gather} 

\noindent The discrete flow rule can be simplified by application of \eqref{eq:disc_eqps_evolution}, \eqref{eq:disc_kin_internal}, and the identity \\ $\left(\bs{G}^p\right)^n = \left( \bar{\bs{F}}^{n} \right)^{-1} \left(\bebar \right)^n \left( \bar{\bs{F}}^{n} \right)^{-T}$:

\begin{gather}
\begin{aligned}
\text{dev} \left[ \bar{\bs{F}}^n \left( \bs{G}^p \right)^{n}  \left( \bar{\bs{F}}^n \right)^T \right] -  \text{dev} \left[ \bar{\bs{F}}^n  \left( \bs{G}^p \right)^{n-1} \left( \bar{\bs{F}}^n \right)^T \right] &= -\frac{2}{3} \Delta \gamma 
\text{tr} \left[ \left( \bebar \right)^n \right] \bs{n}^n, \\
\left(\bzeta \right)^n - \text{dev} \left[   \left( \bar{\bs{F}}^{n} \right) \left( \bar{\bs{F}}^{n-1} \right)^{-1} \left(\bebar \right)^{n-1} \left( \bar{\bs{F}}^{n-1} \right)^{-T}  \left( \bar{\bs{F}}^n \right)^T \right]  &= -2 \Delta \gamma \left( \Ie \right)^n \bs{n}^n, \\
\left(\bzeta \right)^n -  \text{dev} \left[ \left(\bebar \right)^n_{\text{trial}}  \right] &= -2 \sqrt{\frac32} \left( \alpha^n - \alpha^{n-1} \right)  \left( \Ie \right)^n \bs{n}^n. 
\end{aligned} \label{eq:disc_flow_rule}
\end{gather}

The other equations in $\bs{C}^n$ are straightforward to discretize.  The discrete constitutive model evolution equations for a load step with plastic flow are

\begin{gather}
\begin{cases}
\begin{aligned}
\left(\bzeta \right)^n -  \text{dev} \left[ \left(\bebar\right)^n_{\text{trial}}  \right] + 2 \sqrt{\frac{3}{2}} \left( \alpha^{n} - \alpha^{n-1} \right)  \left(\Ie \right)^n  \frac{\bs{s}^n}{\| \bs{s}^n \|}   &= \bs{0}, \\[8pt]
\text{det} \left[ \left(\bzeta \right)^n + (\Ie)^{n} \bs{I}  \right] &= 1, \\[2pt]
\| \bs{s}^n \| -  \sqrt{\frac{2}{3}} \left[ Y + K \alpha^n + S \left\lbrace 1 - \exp \left(-D \alpha^n \right) \right\rbrace \right] &= 0.
\end{aligned}
\end{cases}\label{eq:Cn_plastic}
\end{gather}

\noindent The local state variable $\bzeta$ is symmetric, so $\bs{C}^n$ contains a total of eight independent equations.

We note that various alternative formulations of this model are well-known in the computational plasticity literature. For example, \cite{de2011computational} describes a formulation based on logarithmic strain measures that makes use of spectral decompositions and features a single equation return mapping algorithm. Such a formulation is amenable to our AD-based approach, but the definition (and number) of local state variables $\bs{\xi}$ would change as would the definitions of $\bs{R}^n$ and $\bs{C}^n$. Lastly, in general for a given constitutive model the definitions of the local state variables and residuals are not unique (see \cite{alberdi2018unified} for a detailed discussion of this point).

\subsection{Linearization and Solution Strategy}

The discrete forward model can be represented as a sum over load steps of coupled residuals:

\begin{gather}
\begin{cases}
\bs{R}^n\left( \bs{U}^n, \bs{\xi}^{n}, \bs{\beta} \right) = 0 \quad &n = 1, \ldots, N_L, \\
\bs{C}^n_e\left( \bs{U}^n_e,  \bs{U}^{n-1}_e, \bs{\xi}^{n}_e, \bs{\xi}^{n-1}_e, \bs{\beta} \right) = 0  \quad &e = 1, \ldots, n_{el}, \ n = 1, \ldots, N_L, \label{eq:disc_forward_model}
\end{cases}
\end{gather}

\noindent where we have included the dependence on the material parameter vector $\bbeta$. The subscript $e$ in $\bs{C}^n_e$ is written here to emphasize that a distinct equation $\bs{C}^n_e$ is solved in each element. We can use an element index here because each element contains a single Gauss point that involves the $\bs{\xi}^n_e$ variable, but if there were more we would also need require an index for them. 

In subsequent text, we make reference to \emph{gathering} and \emph{scattering}
variables. Gathering refers to collecting appropriate element-level data from a global linear
algebra data structure (i.e.\ a data structure defined over the entire mesh). For instance,
gathering $\bs{U}^n_e$ would refer to collecting the nodal displacement and pressure data
associated with element $\mathcal{B}_e$ from the
global state vector $\bs{U}^n$ at load step $n$. Conversely, scattering refers to
contributing (either via assignment or summation) of local element-level data into
a global linear algebra data structure.

Our element-level solution procedure involves gathering the current and previous values of global and local state variables for a given element. We first solve the system represented by $\bs{C}^n_e = \bzero$ for $\bs{\xi}^n_e$ by \emph{seeding} the AD objects that represent the local state variables. More specifically, each $\bs{\xi}^n_e$ contains a scalar value and derivative array of length 8 (one for each component of $\bs{\xi}^n_e$). The seeding operation sets the entry of derivative array that corresponds to the local state variable of interest to 1 and all other entries to zero. As AD objects are combined using elementary operations and/or are composed with elementary functions, operator overloading as implemented by \texttt{Sacado} updates the values and derivative arrays according to the chain rule such that the final product has the correct values and derivatives. Thus after the local state variables have been seeded, all that is required to compute the matrix for the Newton solve $\frac{\pt \bs{C}^n_e}{\pt \bs{\xi}^n_e}$ is to evaluate $\bs{C}^n_e$ in terms of these AD objects. We note that the global state variables $\bs{U}^n_e$ and parameters $\bs{\beta}$ must be unseeded for this procedure to yield the correct result (i.e.\ their values are set from the gather operation but their derivative arrays are empty).

Once the values of $\bs{\xi}^n_e$ has been determined from this nonlinear solve, the next step is to determine the element-level tangent stiffness matrix for the global Newton solve of the system $\bs{R}^n = \bs{0}$

\begin{gather}
 \frac{d \bs{R}^n_e}{d \bs{U}^n_e} = \frac{\partial \bs{R}^n_e}{\partial \bs{U}^n_e} + \frac{\partial \bs{R}^n_e}{\partial \bs{\xi}^n_e} \frac{\partial \bs{\xi}^n_e}{\partial \bs{U}^n_e}. \label{eq:global_tangent_stiffness}
\end{gather}

\noindent We note that to compute \eqref{eq:global_tangent_stiffness} the sensitivities of the local state variables with respect to global state variables $\frac{\pt \bs{\xi}^n_e}{\pt \bs{U}^n_e}$ are needed. They are expressed through the following relationship
\begin{gather}
\frac{\partial \bs{C}^n_e}{\partial \bs{\xi}^n_e} \frac{\partial \bs{\xi}^n_e}{\partial \bs{U}^n_e} = -\frac{\partial \bs{C}^n_e}{\partial \bs{U}^n_e},
\label{eq:local_res_forward_sens}
\end{gather}
\noindent which is obtained by performing a forward sensitivity analysis on $\bs{C}^n_e$. The right hand side of \eqref{eq:local_res_forward_sens} is obtained by unseeding the local state variables, seeding the global state variables $\bs{U}^n_e$, and then evaluating $\bs{C}^n_e$ once more (its derivative array will contain the entries of $\frac{\partial \bs{C}^n_e}{\partial \bs{U}^n_e}$). The left hand side can be reused from the Newton solve for $\bs{C}^n_e$.

We then populate the derivative array of $\bs{\xi}^n_e$ with the entries of the $\frac{\partial \bs{\xi}^n_e}{\partial \bs{U}^n_e}$ while also seeding $\bs{U}^n_e$ and evaluate $\bs{R}^n_e$, which also produces $\frac{d \bs{R}^n_e}{d \bs{U}^n_e}$ through the use of AD. These element-level quantities are scattered to a global linear system that is solved to obtain the Newton update to $\bs{U}^n$. The process continues until both the global equilibrium and local constitutive equation residuals have converged as determined by sufficiently ``tight'' Newton solve tolerances $\epsR$ and $\epsC$. This procedure is described in detail in Algorithm \ref{alg:forward}.

\begin{algorithm}
\caption{Forward Problem}
\begin{spacing}{1.5}
\begin{algorithmic}
\Require $\epsR > 0$, \; $\epsC > 0$ \Comment{residual tolerances}
\State $\bs{U}^0 \leftarrow \bzero$, \;
$\left(\bzeta\right)^0 \leftarrow \bzero$, \;
$\left(\Ie\right)^0 \leftarrow 1$, \;
$\alpha^0 \leftarrow 0$ \Comment{initialize global and local state variables}
\For{$n=1,\dots,N_L$} \Comment{loop over all load steps}
\State $\bs{U}^n \leftarrow \bs{U}^{n-1}$, \;
$\bs{\xi}^n \leftarrow \bs{\xi}^{n-1}$ \Comment{initialize global and local state variables}
\While{$\| \bs{R}^n(\bs{U}^n, \bs{\xi}^n) \| > \epsR$} \Comment{global residual convergence criteria}
\For{$e = 1, \dots, n_{el}$} \Comment{loop over all elements}
\State Gather $\bs{U}^n_e, \bs{U}^{n-1}_e, \bs{\xi}^{n}_e$, and $ \bs{\xi}^{n-1}_e$ \Comment{gather local element variables}
\State Compute $\left(f^n_e\right)_{\text{trial}}$ to determine the form of $\bs{C}^n_e$ \Comment{check for plastic deformation}
\While{$\| \bs{C}^n_e(\bs{U}^n_e, \bs{U}^{n-1}_e, \bs{\xi}^n_e, \bs{\xi}^{n-1}_e) \| > \epsC$} \Comment{local residual convergence criteria}
\State Seed $\bs{\xi}^n_e$ and evaluate $\bs{C}^n_e$ to obtain $\frac{\partial\bs{C}^n_e}{\partial \bs{\xi}^n_e}$ \Comment{obtain derivatives with AD}
\State Solve $\frac{\partial \bs{C}^n_e}{\partial \bs{\xi}^n_e} \Delta \bs{\xi}_e = -\bs{C}^n_e$ \Comment{solve for local nonlinear update}
\State $\bs{\xi}^n_e \leftarrow \bs{\xi}^n_e + \Delta \bs{\xi}_e$ \Comment{update the local state variables}
\EndWhile
\State Unseed $\bs{\xi}^n_e$, seed $\bs{U}^n_e$, and evaluate $\bs{C}^n_e$ to obtain $\frac{\partial \bs{C}^n_e}{\partial \bs{U}^n_e}$ \Comment{obtain derivatives with AD}
\State Solve $\frac{\partial \bs{C}^n_e}{\partial \bs{\xi}^n_e} \frac{\partial \bs{\xi}^n_e}{\partial \bs{U}^n_e} = -\frac{\partial \bs{C}^n_e}{\partial \bs{U}^n_e}$ \Comment{obtain derivatives with a linear solve}
\State Seed $\bs{\xi}^n_e$ with $\frac{\partial \bs{\xi}^n}{\partial \bs{U}^n_e}$ \Comment{seed local state variables with derivative information}
\State Evaluate $\bs{R}^n_e$ to obtain $\frac{d \bs{R}^n_e}{d \bs{U}^n_e} = \frac{\partial \bs{R}^n_e}{\partial \bs{U}^n_e} + \frac{\partial \bs{R}^n_e}{\partial \bs{\xi}^n_e} \frac{\partial \bs{\xi}^n_e}{\partial \bs{U}^n_e}$ \Comment{obtain total derivative with AD}
\State Scatter $\bs{R}^n_e$ and $\frac{d \bs{R}^n_e}{d \bs{U}^n_e}$ \Comment{scatter element contributions to global data}
\EndFor
\State Solve $\frac{d \bs{R}^n}{d \bs{U}^n} \Delta \bs{U} = -\bs{R}^n$ \Comment{solve for global nonlinear update}
\State $\bs{U}^n \leftarrow \bs{U}^n + \Delta \bs{U}$ \Comment{update the global state variables}
\EndWhile
\EndFor
\end{algorithmic}
\end{spacing}
\label{alg:forward}
\end{algorithm}

The MPI-parallelized software implementation of \eqref{eq:disc_forward_model} is written in \texttt{C++}. It makes use of several packages in the Trilinos project: \texttt{Sacado} for forward mode AD through expression templates, \texttt{Tpetra} for distributed linear algebra objects, \texttt{MueLu} for algebraic multigrid preconditioning, \texttt{Belos} for matrix-free iterative linear solvers, and the \texttt{MiniTensor} package for convenient storage and manipulation of tensor objects \cite{trilinos-website}. We also utilize the software \texttt{SCOREC} (\url{https://github.com/SCOREC/core}) for mesh data-structures, I/O, and discretization-related computations.

\section{Inverse Formulations}
\label{sec:inverse}

Now that we have discussed the forward problem and its solution, we introduce the
objective function for the full-field model calibration problem. We then discuss three means of obtaining the objective function gradient including a FEMU approach that uses finite difference approximations and forward and adjoint sensitivities-based approaches. As stated in the introduction, our presentation of the equations obtained through forward and adjoint local sensitivity analyses of the coupled residuals follows that taken in \cite{michaleris1994tangent,kulkarni2007newton,alberdi2018unified}. A primary contribution of our work is the application of AD to these methods to obtain the derivatives of the residuals with respect to the global and local state variables and parameters.

The objective functions employed for constitutive model calibration with experimental measurements are more complicated in practice than the ones in our presentation, and we have omitted these details to simplify the optimization formulations and numerical examples. First, we assume that the BCs of the FE model are known exactly. In practice a force-matching term that incorporates load frame data in the form of a resultant force vector is added to the objective function, as the traction BCs imposed by the testing apparatus are typically unknown and force information is needed to calibrate some parameters (e.g.\ Young's modulus). Second, it is also common to introduce weighting functions that weight both the displacement and force-matching terms according to the uncertainty in their respective measurements \cite{mahnken1997parameter, mahnken1999aspects, mahnken2000comprehensive, mahnken2002theoretical, neggers2017improving, neggers2017reducing, neggers2018big, neggers2019simultaneous}.

\subsection{Constrained Optimization Formulation}

The objective function is represented as a surface integral of the displacement mismatch over the initially-planar surface where DIC data $\bs{d}$ is available:

\begin{gather}
\begin{aligned}
\underset{\bs{\beta}}{\text{min}} \quad &\mathcal{J} :=  \sum_{n=1}^{N_L}\mathcal{J}^n = \sum_{n=1}^{N_L}  \frac12 \int_{\Gamma_{\text{DIC}}} || \bs{u}^n - \bs{d}^n ||^2 \  dA, \\
\text{s.t.} \quad & \bs{R}^n\left( \bs{U}^n, \bs{\xi}^{n}, \bs{\beta} \right) = 0, \quad n = 1, \ldots, N_L,  \\
& \bs{C}^n_e \left( \bs{U}^n_e,  \bs{U}^{n-1}_e, \bs{\xi}^{n}_e, \bs{\xi}^{n-1}_e, \bs{\beta} \right) = 0, \quad e = 1, \ldots, n_{el}, \ n = 1, \ldots, N_L, \\
& \bs{\beta}_{\text{lo}} \leq \bs{\beta} \leq \bs{\beta}_{\text{up}}. \label{eq:opt_problem}
\end{aligned}
\end{gather}

\noindent The use of piecewise-linear shape functions for $\bs{u}$ and $\bs{d}$ results in a quadratic objective function that can be exactly integrated using a three point Gaussian quadrature scheme for triangles.

 Many formulations of inverse problems are ill-posed in the sense that they fail one of three conditions satisfied by well-posed problems: existence, uniqueness, and stability (``small'' perturbations in the input data lead to ``small'' perturbations in the solution). When the inverse problem solution is a discretized field quantity (e.g.\ a heterogeneous material parameter represented using a finite basis with a dimension commensurate with the number of nodes in the mesh) the quality of the inverse problem solution is often extremely poor without the introduction of a regularization term in the objective function, which is typically in the form of a penalty on a norm or semi-norm of the parameter field (e.g.\ $L^2$, $H^1$, and total variation regularization), and in a fashion serves to transform the ill-posed inverse problem into a ``nearby'' well-posed one.
 
There is a correspondence between the regularization functions employed in deterministic solutions of inverse problems and the prior distributions in non-deterministic Bayesian formulations, such that the use of regularization can be viewed as a means of injecting prior knowledge of parameter values into the inverse problem formulation. The complete absence of regularization terms corresponds to the use of a non-informative prior, while the specification of a prior with independent uniform marginal distributions corresponds to the imposition of box constraints.
    
    In the problems examined in this work (where the constitutive model parameters do not vary with space and consequently $\bs{\beta}$ is low-dimensional) we found that regularization was not needed to obtain accurate solutions, as evidenced by our numerical results in section \ref{sec:results}. We recover close approximations of the model parameters from noise-corrupted data, and the error is proportional to the amount of noise.
    
In \cite{long2015fast} the authors express a similar sentiment regarding their lack of regularization, but they base their argument in terms of the dimension of observation space being large compared to the dimension of the parameter space (for the discretized inverse problem). The problems studied in this work all satisfy this property, and it could heuristically be argued that the class of calibration problems for homogeneous constitutive model parameters from full-field data would as well.

In this work we take the \textit{reduced-space} approach to solving \eqref{eq:opt_problem} wherein the constraints are satisfied at each optimization iterate, which amounts to at least one complete solution of the forward problem per iteration (see \cite{hicken2013comparison} for a detailed comparison between reduced and full space approaches). For minimization we employ the bound-constrained optimization algorithm L-BFGS-B \cite{zhu1997algorithm} with a line search for globalization as implemented by the Rapid Optimization Library (\texttt{ROL}), an optimization software package in the Trilinos project \cite{rol-website}. It requires evaluations of the objective function and its gradient. In subsequent subsections we describe three approaches to obtaining the objective function gradient.

\subsection{Finite Difference Approximation}

The finite difference approach is the most straightforward route to the gradient. It only requires an implementation of the forward model and the ability to evaluate the objective function. The simplicity of this scheme has resulted in its widespread adoption in the literature. Algorithm \ref{alg:fd_gradient} contains the details for the finite difference approximations of the objective function gradient.

\begin{algorithm}
\caption{Finite Difference Gradient}
\begin{spacing}{1.5}
\begin{algorithmic}
\Require $\bs{\beta}$ \Comment{values for the material parameters}
\Require $\varepsilon_{\text{FD}} > 0$ \Comment{finite difference step size}
\State $\bs{u}_{\text{ref}} \leftarrow$ Solve \eqref{eq:disc_forward_model} using $\bs{\beta}$ \Comment{forward model displacements from unperturbed parameters}
\State $\obj_{\text{ref}} \leftarrow \obj (\bs{u}_{\text{ref}})$ \Comment{objective function from unperturbed parameters}
\For{$i = 1, \dots, N_{\bs{\beta}}$} \Comment{loop over all material parameters}
\State $\hat{\bs{\beta}}_i \leftarrow \bs{\beta} + \varepsilon_{\text{FD}} \bs{e}_i$ \Comment{perturb the $i^{th}$ parameter}
\State $\hat{\bs{u}}_i \leftarrow$ Solve \eqref{eq:disc_forward_model} using $\hat{\bs{\beta}}_i$ \Comment{forward model displacements from perturbed parameter}
\State $\hat{\obj}_i \leftarrow \obj (\hat{\bs{u}}_i)$ \Comment{objective function from perturbed parameters}
\State $\left[ \objgrad \right]_i \leftarrow \frac{1}{\varepsilon_{\text{FD}}} \left( \hat{\mathcal{J}}_i  - \mathcal{J}_{\text{ref}} \right)$ \Comment{$i^{th}$ objective gradient component}
\EndFor
\end{algorithmic}
\end{spacing}
\label{alg:fd_gradient}
\end{algorithm}

There are two drawbacks associated with this approach. First, the accuracy of the approximation depends on the value of $\varepsilon_{\text{FD}}$. Large values result in a larger truncation error while small values suffer from cancellation error due to the finite precision of floating point numbers. A ``good'' choice for $\varepsilon_{\text{FD}}$ lies in between these regimes, and in our work we default to the value chosen by \texttt{ROL}.

A second, more significant disadvantage is the computational cost. The forward problem must be solved for each component of the gradient, resulting in a total of $N_{\bs{\beta}} + 1$ nonlinear, pseudo-transient solves per gradient and/or sensitivity matrix calculation where $N_{\bs{\beta}}$ is the number of optimization parameters. Thus, this method scales poorly with the dimension of the parameter space. Increasing the order of the finite difference approximation can improve its accuracy but will also increase the computational cost further.

We note that the semi-analytical methods employed in the structural and shape optimization literature (see \cite{fernandez2018semi} for a review) are based upon the forward and adjoint sensitivity approaches described in the next sections. The difference between these methods and those described in this work is the use of a finite difference approximation of the derivative of the residuals and objective function (when it depends on $\bs{\beta}$) w.r.t.\ the parameters in semi-analytical methods. Consequently, these methods inherit the computational efficiency of forward and adjoint sensitivity formulations but suffer from the round-off or truncation errors that arise from finite difference approximations.

\subsection{Forward Sensitivities}
The second approach to obtaining the gradient is referred to as forward or direct sensitivity analysis. In this approach we treat the objective function as an implicit function of material parameters and differentiate w.r.t $\bbeta$ to obtain

\begin{gather}
\objgrad = \sum_{n=1}^N \left( \frac{\partial \mathcal{J}^n}{\partial \bs{U}^n} \frac{\partial \bs{U}^n}{\partial \bs{\beta}} + \cancel{\frac{\partial \mathcal{J}^n}{\partial \bs{\xi}^n} \frac{\partial \bs{\xi}^n}{\partial \bs{\beta}}} + \cancel{\frac{\partial \mathcal{J}^n}{\partial \bs{\beta}} } \right)= \sum_{n=1}^{N_L} \sum_{e=1}^{n_{el}} \frac{\partial \mathcal{J}^n_e}{\partial \bs{u}^n_e} \frac{\partial \bs{u}^n_e}{\partial \bs{\beta}}.
\end{gather}

\noindent We note that the objective function does not have an explicit dependence on the pressure or local state variables. The quantity $\frac{\partial \mathcal{J}^n}{\partial \bs{u}^n}$ may be obtained analytically for our objective function, but in other applications when it is more complex (e.g.\ when it contains a force-matching term) the use of AD may be warranted. The challenge in the forward sensitivities approach is the determination of $\frac{\partial \bs{u}^n_e}{\partial \bs{\beta}}$. The first step in the derivation is to differentiate the forward problem w.r.t $\bbeta$:

\begin{gather}
\begin{aligned}
 \frac{\pt \bs{R}^n}{ \pt \bs{U}^n} \ufsc  + \sum_{e=1}^{n_{el}} \frac{\pt \bs{R}^n_e}{ \pt \bs{\xi}^n_e} \xifsce +  \frac{\pt \bs{R}^n}{ \pt \bs{\beta}}   &= \bzero,  \quad n = 1, \ldots, N_L, \\
 \frac{\pt \bs{C}^n}{ \pt \bs{U}^n_e} \ufsce + \frac{\pt \bs{C}^n_e}{ \pt \bs{\xi}^n_e} \xifsce
+ \frac{\pt \bs{C}^n_e}{ \pt \bs{\beta}}  +  \frac{\pt \bs{C}^n_e}{ \pt \bs{U}^{n-1}_e} \ufspe + \frac{\pt \bs{C}^n_e}{ \pt \bs{\xi}^{n-1}_e} \xifspe  &= \bzero, \quad e = 1, \ldots, n_{el}, \ n = 1, \ldots, N_L.
\end{aligned} \label{eq:fs_full}
\end{gather}

We have a block system of linear equations that contains the matrices $A^n_e :=  \frac{\pt \bs{R}^n_e}{ \pt \bs{U}^n_e}$, $B^n_e :=  \frac{\pt \bs{R}^n_e}{ \pt \bs{\xi}^n_e}$, $C^n_e :=  \frac{\pt \bs{C}^n_e}{ \pt \bs{U}^n_e}$, and $D^n_e := \frac{\pt \bs{C}^n_e}{ \pt \bs{\xi}^n_e}$  with matrix unknowns $X^n := \frac{\pt \bs{U}^n}{\pt \bs{\beta}}$ and $Y^n_e := \xifsce$:

\begin{gather}
\begin{aligned}
& \sum_{e=1}^{n_{el}} \left(A^n_e \right) X^n + \sum_{e=1}^{n_{el}} B^n_e Y^n_e = F^n, &&n = 1, \ldots, N_L,  \\
&C^n_e X^n_e+ D^n_e Y^n_e = G^n_e, &&n = 1, \ldots, N_L,  \quad e = 1, \ldots, n_{el}, \\
&F^n := -  \sum_{e=1}^{n_{el}} \frac{\pt \bs{R}^n_e}{ \pt \bs{\beta}}, \\
&G^n_e := - \left( \frac{\pt \bs{C}^n_e}{ \pt \bs{\beta}}  +  \frac{\pt \bs{C}^n_e}{ \pt \bs{U}^{n-1}_e} \ufspe +  \frac{\pt \bs{C}^n_e}{ \pt \bs{\xi}^{n-1}_e} \xifspe \right), \label{eq:fs_linear_system}
\end{aligned}
\end{gather}

\noindent where $X^n_e$ is the element-level (i.e.\ gathered) representation of $X^n$.

Our use of AD allows us to obtain every quantity in \eqref{eq:fs_linear_system} through element-level calculations where we seed one of $\bs{U}^n_e, \bs{U}^{n-1}_e, \bs{\xi}^n_e, \bs{\xi}^{n-1}_e$, or $\bs{\beta}$, evaluate $\bs{R}^n_e$ or $\bs{C}^n_e$, and compute the necessary quantities for the various linear systems. The global linear system for $X^n$ is assembled and solved first, and then $Y^n_e$ is computed in a second pass for each element following a Schur complement approach

\begin{gather}
\begin{aligned}
&\sum_{e=1}^{n_{el}}  \left(A^n_e - B^n_e \left(D^n_e \right)^{-1} C^n_e \right) X^n = F^n -  \sum_{e=1}^{n_{el}} \left( B^n_e \left( D^n_e \right)^{-1} G^n_e \right), &&n = 1, \ldots, N_L,  \\
&Y^n_e = \left(D^n_e \right)^{-1} \left( G^n_e - C^n_e X^n_e \right), &&n = 1, \ldots, N_L,  \quad e = 1, \ldots, n_{el}. \label{eq:schur_fs}
\end{aligned}
\end{gather}

\noindent We note that $G^n_e$, an element-level quantity, is needed for both systems so we store it in a global object and overwrite it during subsequent load steps. Algorithm \ref{alg:forward_sens_gradient} contains a complete description of the forward sensitivities approach to obtaining the gradient.

\begin{algorithm}
\caption{Gradient from Forward Sensitivities}
\begin{spacing}{1.5}
\begin{algorithmic}
\Require $\varepsilon_{\alpha} > 0$ \Comment{plastic deformation tolerance}
\State $\frac{\pt\bs{U}^0}{\pt\bs{\beta}} \leftarrow  \bzero$, \;
$\frac{\pt\bs{\xi}^0}{\pt\bs{\beta}} \leftarrow  \bzero$, \;
$\frac{\pt \mathcal{J}}{\pt \bs{\beta}} \leftarrow  \bzero$ \Comment{initialize sensitivity matrices and gradient}
\For{$n=1,\dots,N_L$} \Comment{loop over all load steps}
\For{$e = 1, \dots, n_{el}$} \Comment{loop over all elements}
\State Gather $\bs{U}^n_e,  \bs{U}^{n-1}_e, \bs{\xi}^{n}_e, \bs{\xi}^{n-1}_e, \frac{\pt\bs{U}^{n-1}_e}{\pt\bs{\beta}}$, and $\frac{\pt\bs{\xi}^{n-1}_e}{\pt\bs{\beta}}$ \Comment{gather local element variables}
\State Seed $\bs{U}^n_e$ and evaluate $\bs{R}^n_e$ to obtain $\frac{\partial \bs{R}^n_e}{\partial \bs{U}^n_e}$ \Comment{obtain derivatives with AD}
\State Compute $\alpha^n_e - \alpha^{n-1}_e > \varepsilon_{\alpha}$ to determine the form of $\bs{C}^n_e$ \Comment{check for plastic deformation}
\State Evaluate $\bs{C}^n_e$ to obtain  $\frac{\partial \bs{C}^n_e}{\partial \bs{U}^n_e}$ \Comment{obtain derivatives with AD}
\State Unseed $\bs{U}^n_e$, seed $\bs{U}^{n-1}_e$, and evaluate $\bs{C}^n_e$ to obtain $\frac{\partial \bs{C}^n_e}{\partial \bs{U}^{n-1}_e}$  \Comment{obtain derivatives with AD}
\State Unseed $\bs{U}^{n-1}_e$, seed $\bs{\xi}^{n-1}_e$, and evaluate $\bs{C}^n_e$ to obtain $\frac{\partial \bs{C}^n_e}{\partial \bs{\xi}^{n-1}_e}$ \Comment{obtain derivatives with AD}
\State Unseed $\bs{\xi}^{n-1}_e$, seed $\bs{\beta}$, and evaluate $\bs{C}^n_e$ to obtain $\frac{\partial \bs{C}^n_e}{\partial \bs{\beta}}$ \Comment{obtain derivatives with AD}
\State Compute $G^n_e = - \left( \frac{\pt \bs{C}^n_e}{ \pt \bs{\beta}}  +  \frac{\pt \bs{C}^n_e}{ \pt \bs{U}^{n-1}_e} \ufspe +  \frac{\pt \bs{C}^n_e}{ \pt \bs{\xi}^{n-1}_e} \xifspe \right)$ \Comment{compute local RHS matrix}
\State Evaluate $\bs{R}^n_e$ to obtain  $\frac{\partial \bs{R}^n_e}{\partial \bs{\beta}}$  \Comment{obtain derivatives with AD}
\State Unseed $\bs{\beta}$, seed $\bs{\xi}^n_e$, and evaluate $\bs{R}^n_e$ to obtain $\frac{\partial \bs{R}^n_e}{\partial \bs{\xi}^n_e}$ \Comment{obtain derivatives with AD}
\State Evaluate $\bs{C}^n_e$ to obtain $\frac{\partial \bs{C}^n_e}{\partial \bs{\xi}^n_e}$ \Comment{obtain derivatives with AD}
\State Scatter $\frac{\partial \bs{R}^n_e}{\partial \bs{U}^n_e} - \frac{\partial \bs{R}^n_e}{\partial \bs{\xi}^n_e} \left( \frac{\partial \bs{C}^n_e}{\partial \bs{\xi}^n_e} \right)^{-1} \frac{\partial \bs{C}^n_e}{\partial \bs{U}^n_e}$ to the LHS \Comment{scatter element contributions to global data}
\State Scatter $-\frac{\partial \bs{R}^n_e}{\partial \bs{\beta}} - \frac{\partial \bs{R}^n_e}{\partial \bs{\xi}^n_e} \left( \frac{\partial \bs{C}^n_e}{\partial \bs{\xi}^n_e} \right)^{-1} G^n_e$ to the RHS \Comment{scatter element contributions to global data}
\State Scatter $G^n_e$ \Comment{scatter element contributions to global data}
\EndFor
\State Solve the global linear system to obtain $\frac{\pt\bs{U}^n}{\pt\bs{\beta}}$ \Comment{solve for global sensitivity matrices}
\For{$e = 1, \dots, n_{el}$} \Comment{loop over all elements}
\State Gather $\bs{U}^n_e,  \bs{U}^{n-1}_e, \bs{\xi}^{n}_e, \bs{\xi}^{n-1}_e, \frac{\pt\bs{U}^{n}_e}{\pt\bs{\beta}}$, and $G^n_e$ \Comment{gather local element variables}
\State Seed $\bs{U}^n_e$ and evaluate $\mathcal{J}^n_e$ to obtain $\frac{\partial \mathcal{J}^n_e}{\partial \bs{U}^n_e}$ \Comment{obtain derivatives with AD}
\State Compute $\alpha^n_e - \alpha^{n-1}_e > \varepsilon_{\alpha}$ to determine the form of $\bs{C}^n_e$ \Comment{check for plastic deformation}
\State Evaluate $\bs{C}^n_e$ to obtain $\frac{\partial \bs{C}^n_e}{\partial \bs{U}^n_e}$ \Comment{obtain derivatives with AD}
\State Unseed $\bs{U}^n_e$, seed $\bs{\xi}^{n}_e$, and evaluate $\bs{C}^n_e$ to obtain $\frac{\partial \bs{C}^n_e}{\partial \bs{\xi}^{n}_e}$ \Comment{obtain derivatives with AD}
\State Solve $\frac{\partial \bs{C}^n_e}{\partial \bs{\xi}^{n}_e} \frac{\pt\bs{\xi}^n_e}{\pt\bs{\beta}} = G^n_e - \frac{\partial \bs{C}^n_e}{\partial \bs{U}^n_e}\frac{\pt\bs{U}^{n}_e}{\pt\bs{\beta}}$ \Comment{solve for local sensitivity matrices}
\State $\objgrad \leftarrow \objgrad + \frac{\partial \mathcal{J}^n_e}{\partial \bs{U}^n_e} \frac{\pt\bs{U}^{n}_e}{\pt\bs{\beta}}$ \Comment{accumulate gradient}
\State Scatter $\frac{\pt\bs{\xi}^n_e}{\pt\bs{\beta}}$ \Comment{scatter element contributions to global data}
\EndFor
\EndFor
\end{algorithmic}
\end{spacing}
\label{alg:forward_sens_gradient}
\end{algorithm}

An attractive property of the forward sensitivities method is that it can be executed during the forward problem solve by appropriately combing Algorithms \ref{alg:forward} and \ref{alg:forward_sens_gradient}. It also only requires the storage of global and local state variables and their associated sensitivity matrices from two neighboring load steps. We also note that the linear solver tolerances for the global and local solves should be set to ``tight'' values and this also goes for the linear solves required for the adjoint approach described in the next section.  Lastly, there are large-scale iterative linear solvers (e.g.\ block GMRES) designed for problems with multiple right hand sides like this, and their application is another potential avenue for improving the computational efficiency of this approach. 

The cost of this method, however, still scales with the dimension of the parameter space like the finite difference approach, but it costs less because it requires the solution of linear systems instead of nonlinear ones. Fortunately, the dimension of the parameter space in the calibration of elastoplastic constitutive models is typically low compared to the number of degrees of freedom in the forward problem. In the next section we will present an approach with a cost that is independent of the dimension of the parameter space, and is typically the only viable option when the dimension of the parameter space is commensurate with the dimension of the state space.

\subsection{Adjoint Sensitivities}

The third approach to computing the gradient is known as the adjoint method. One way to derive it is through the use of the Lagrangian:

\begin{gather}
\mathcal{L}\left(\bs{U}, \bs{\xi}, \bs{\eta},  \bs{\phi}, \bs{\beta} \right) =  \sum_{n=1}^{N_{L}}  \left( \mathcal{J}^n(\bs{U}^n) +  (\bs{\eta}^n)^T \bs{R}^n( \bs{U}^n, \bs{\xi}^{n}, \bs{\beta} ) + \sum_{e=1}^{N_{el}}  (\bs{\phi}^n_e)^T \bs{C}^n_e( \bs{U}^n_e,  \bs{U}^{n-1}_e, \bs{\xi}^{n}_e, \bs{\xi}^{n-1}_e, \bs{\beta} ) \right),
\end{gather}

\noindent where the variables $\bs{\eta}^n$ and $\bs{\phi}^n_e$ are Lagrange multipliers for the equilibrium PDE (global) and constitutive equation residuals (local), respectively. The constraint (also called state) equations are recovered when the Lagrangian is differentiated with respect to these multipliers. The \textit{adjoint} (also known as co-state) equations are obtained by differentiating the Lagrangian with respect to the state variables $\bs{U}^n$ and $\bs{\xi}^n_e$ and transposing the result, 

\begin{gather}
\begin{aligned}
&\left( \frac{ \pt \mathcal{J}^{N_L}}{\pt \bs{U}^{N_L}} \right)^T + \left(\frac{\pt \bs{R}^{N_L}}{ \pt \bs{U}^{N_L}}\right)^T \bs{\eta}^{N_L} + \sum_{e=1}^{n_{el}} \left(\frac{\pt \bs{C}^{N_L}_e}{ \pt \bs{U}^{N_L}_e}\right)^T \bs{\phi}^{N_L}_e = \bzero, \\
&\left( \frac{\pt \bs{R}^{N_L}_e}{ \pt \bs{\xi}^{N_L}_e} \right)^T \bs{\eta}^{N_L}_e + \left(\frac{\pt \bs{C}^{N_L}_e}{ \pt \bs{\xi}^{N_L}_e} \right)^T  \bs{\phi}^{N_L}_e = \bzero, \quad e = 1, \ldots, n_{el}, \\
&\left( \frac{ \pt \mathcal{J}^n}{\pt \bs{U}^n} \right)^T + \left(\frac{\pt \bs{R}^n}{ \pt \bs{U}^n}\right)^T \bs{\eta}^n + \sum_{e=1}^{n_{el}}  \left( \left(\frac{\pt \bs{C}^n_e}{ \pt \bs{U}^n_e}\right)^T \bs{\phi}^n_e +  \left(\frac{\pt \bs{C}^{n+1}_e}{ \pt \bs{U}^n_e}\right)^T \bs{\phi}^{n+1}_e \right) = \bzero,  \quad n = N_L - 1, \ldots, 1, \\
& \left( \frac{\pt \bs{R}^n_e}{ \pt \bs{\xi}^n_e} \right)^T \bs{\eta}^n_e + \left(\frac{\pt \bs{C}^n_e}{ \pt \bs{\xi}^n_e} \right)^T  \bs{\phi}^n_e + \left(\frac{\pt \bs{C}^{n+1}_e}{ \pt \bs{\xi}^n_e} \right)^T  \bs{\phi}^{n+1}_e = \bzero , \quad e = 1, \ldots, n_{el}, \ n = N_L - 1, \ldots, 1.
\end{aligned}
\end{gather}

\noindent We note that unlike the forward sensitivities system \eqref{eq:fs_full}, the adjoint problem is a terminal linear boundary value problem, i.e.\ \emph{it is solved backwards in time}.

As in the forward sensitivities section, we can simplify the notation by introducing the coupled linear systems: 

\begin{gather}
\begin{aligned}
& \sum_{e=1}^{n_{el}} \left( \left(A^{N_L}_e \right)^T \right) x^{N_L} + \sum_{e=1}^{n_{el}} \left(C^{N_L}_e \right)^T y^{N_L}_e = \bar{f}^{N_L},    \\
& \left( B^{N_L}_e \right)^T x^{N_L}_e +  \left(D^{N_L}_e \right)^T y^{N_L}_e = \bzero,  \quad && e = 1, \ldots, n_{el},    \\
&\sum_{e=1}^{n_{el}} \left( \left(A^n_e \right)^T \right) x^n + \sum_{e=1}^{n_{el}} \left(C^n_e \right)^T y^n_e = f^n, &&n = N_L - 1, \ldots, 1,   \\
&\left( B^n_e \right)^T x^n_e +  \left(D^n_e \right)^T y^n_e = g^n_e, &&n = N_L - 1, \ldots, 1,  \quad e = 1, \ldots, n_{el},    \\
\label{eq:adjoint_linear_system}
\end{aligned}
\end{gather}

\noindent with the right hand side vectors

\begin{gather}
\begin{aligned}
& \bar{f}^n := - \sum_{e=1}^{n_{el}} \left(  \left( \frac{ \pt \mathcal{J}^n_e}{\pt \bs{U}^n_e} \right)^T \right) , &&n = N_L, \ldots, 1, \\
& f^n :=  \bar{f}^n + \tilde{f}^n  , &&n = N_L - 1, \ldots, 1, \\
&\tilde{f}^n := - \sum_{e=1}^{n_{el}}  \left(\frac{\pt \bs{C}^{n+1}_e}{ \pt \bs{U}^n_e} \right)^T \bs{\phi}^{n+1}_e, &&n = N_L - 1, \ldots, 1, \\
&g^n_e :=  - \left(\frac{\pt \bs{C}^{n+1}_e}{ \pt \bs{\xi}^n_e} \right)^T  \bs{\phi}^{n+1}_e, &&n = N_L - 1, \ldots, 1, \label{eq:adjoint_linear_system_rhs}
\end{aligned}
\end{gather}

\noindent where $x^n := \bs{\eta}^n$, $x^n_e := \bs{\eta}^n_e$  and $y^n_e := \bs{\phi}^n_e$  and the matrices on the left hand side are the same as those defined in the previous section, such that the linear system is the transpose of the one in the forward sensitivities problem.  It requires knowing the forward solution at each point in time, so it must be stored or generated as needed through checkpointing \cite{cyr2015towards}.

As in the previous section, we solve for the adjoint variables using a two pass approach based on the Schur complement. In this problem
we must compute and store the ``history'' vectors $\tilde{f}^n$ (global) and $g^n_e$ (local). After performing block elimination the coupled linear can be compactly written when these vectors are initialized to $\bs{0}$ at $n = N_L$ as described in Algorithm \ref{alg:adjoint_gradient}.

\begin{gather}
\begin{aligned}
&\sum_{e=1}^{n_{el}} \left( \left(A^n_e \right)^T - \left(C^n_e \right)^T \left(D^n_e \right)^{-T} \left(B^n_e \right)^T \right) x^n = f^n - \sum_{e=1}^{n_{el}}  \left(C^n_e \right)^T \left(D^n_e \right)^{-T} g^n_e, \quad n = N_L, \ldots, 1,   \\
&y^n_e = \left(D^n_e \right)^{-T} \left( g^n_e - \left(B^n_e \right)^T x^n_e \right), \quad n = N_L, \ldots, 1,  \quad e = 1, \ldots, n_{el}.  \label{eq:schur_adjoint}
\end{aligned}
\end{gather}

The expression for the gradient is obtained by differentiating the Lagrangian with respect to $\bs{\beta}$. The contribution to the gradient from each load step can be computed at the element level after the adjoint solutions for that step have been determined and accumulated to obtain the gradient.

\begin{gather}
\frac{\pt \mathcal{L}}{ \pt \bs{\beta}} = \objgrad = \sum_{n=1}^{N_L} \sum_{e=1}^{n_{el}}  \left( \cancel{\frac{\pt \mathcal{J}^n_e}{\pt \bs{\beta}}}  + (\bs{\eta}^n_e)^T \frac{\pt \bs{R}^n_e}{\pt \bs{\beta}} +  (\bs{\phi}^n_e)^T \frac{\pt \bs{C}^n_e}{\pt \bs{\beta}}  \right).
\end{gather}

\begin{algorithm}
\caption{Gradient from Adjoint Sensitivities}
\begin{spacing}{1.5}
\begin{algorithmic}
\Require $\varepsilon_{\alpha} > 0$ \Comment{plastic deformation tolerance}
\State $\tilde{f}^{N_L} \leftarrow \bzero$,\;
$g^{N_L} \leftarrow \bzero$,\;
$\frac{\pt \mathcal{J}}{\pt \bs{\beta}} \leftarrow \bzero$ \Comment{initialize adjoint history vectors and gradient}
\For{$n = N_L, \ldots, 1$} \Comment{loop over all load steps}
\For{$e = 1, \dots, n_{el}$} \Comment{loop over all elements}
\State Gather $\bs{U}^n_e,  \bs{U}^{n-1}_e, \bs{\xi}^{n}_e, \bs{\xi}^{n-1}_e, \tilde{f}^n_e$,  and $g^{n}_e$ \Comment{gather local element variables}
\State Compute $\alpha^n_e - \alpha^{n-1}_e > \varepsilon_{\alpha}$ to determine the form of $\bs{C}^n_e$ \Comment{check for plastic deformation}
\State Seed $\bs{\xi}^n_e$ and evaluate $\bs{C}^n_e$ to obtain $\left(\frac{\partial \bs{C}^n_e}{\partial \bs{\xi}^n_e}\right)^T$ \Comment{obtain derivatives with AD}
\State Evaluate $\bs{R}^n_e$ to obtain $\left(\frac{\partial \bs{R}^n_e}{\partial \bs{\xi}^n_e}\right)^T$ \Comment{obtain derivatives with AD}
\State Unseed $\bs{\xi}^n_e$, seed $\bs{U}^n_e$, and evaluate $\bs{R}^n_e$ to obtain $\left(\frac{\partial \bs{R}^n_e}{\partial \bs{U}^n_e}\right)^T$ \Comment{obtain derivatives with AD}
\State Evaluate $\bs{C}^n_e$ to obtain  $\left(\frac{\partial \bs{C}^n_e}{\partial \bs{U}^n_e}\right)^T$ \Comment{obtain derivatives with AD}
\State Evaluate $\mathcal{J}^n_e$ to obtain $\left( \frac{\partial \mathcal{J}^n_e}{\partial \bs{U}^n_e} \right)^T$ \Comment{obtain derivatives with AD}
\State Scatter $\left( \frac{\partial \bs{R}^n_e}{\partial \bs{U}^n_e} \right)^T - \left(\frac{\partial \bs{C}^n_e}{\partial \bs{U}^n_e} \right)^T\left( \frac{\partial \bs{C}^n_e}{\partial \bs{\xi}^n_e} \right)^{-T} \left(\frac{\partial \bs{R}^n_e}{\partial \bs{\xi}^n_e} \right)^T$ to the global LHS \Comment{scatter local to global}
\State Scatter $-\left(\frac{\partial \mathcal{J}^n_e}{\partial \bs{U}^n_e}\right)^T + \tilde{f}^{n}_e - \left(\frac{\partial \bs{C}^n_e}{\partial \bs{U}^n_e} \right)^T \left( \frac{\partial \bs{C}^n_e}{\partial \bs{\xi}^n_e} \right)^{-T} g^n_e $ to the global RHS \Comment{scatter local to global}
\EndFor
\State Solve the global linear system to obtain $\bs{\eta}^n$ \Comment{solve for global adjoint vector}
\For{$e = 1, \dots, n_{el}$} \Comment{loop over all elements}
\State Gather $\bs{U}^n_e,  \bs{U}^{n-1}_e, \bs{\xi}^{n}_e, \bs{\xi}^{n-1}_e, \bs{\eta}^n_e$, and $g^n_e$ \Comment{gather local element variables}
\State Seed $\bs{\xi}^{n}_e$ and evaluate $\bs{R}^n_e$ to obtain $\left(\frac{\partial \bs{R}^n_e}{\partial \bs{\xi}^{n}_e}\right)^T$ \Comment{obtain derivatives with AD}
\State Compute $\alpha^n_e - \alpha^{n-1}_e > \varepsilon_{\alpha}$ to determine the form of $\bs{C}^n_e$ \Comment{check for plastic deformation}
\State Evaluate $\bs{C}^n_e$ to obtain $\left(\frac{\partial \bs{C}^n_e}{\partial \bs{\xi}^{n}_e}\right)^T$ \Comment{obtain derivatives with AD}
\State Solve $ \left(\frac{\partial \bs{C}^n_e}{\partial \bs{\xi}^{n}_e} \right)^{T}  \bs{\phi}^n_e = g^n_e - \left(\frac{\partial \bs{R}^n_e}{\partial \bs{\xi}^n_e} \right)^T \bs{\eta}^n_e $ \Comment{solve for local adjoint vector}
\State Unseed $\bs{\xi}^{n}_e$, seed $\bs{U}^{n-1}_e$, and evaluate $\bs{C}^n_e$ to obtain $\left(\frac{\partial \bs{C}^n_e}{\partial \bs{U}^{n-1}_e}\right)^T$ \Comment{obtain derivatives with AD}
\State Compute $\tilde{f}^{n-1}_e = -\left(\frac{\partial \bs{C}^n_e}{\partial \bs{U}^{n-1}_e}\right)^T \bs{\phi}^n_e$ \Comment{compute contribution to global history vector}
\State Unseed $\bs{U}^{n-1}_e$, seed $\bs{\xi}^{n-1}_e$, and evaluate $\bs{C}^n_e$ to obtain $\left(\frac{\partial \bs{C}^n_e}{\partial \bs{\xi}^{n-1}_e}\right)^T$ \Comment{obtain derivatives with AD}
\State Compute $g^{n-1}_e = -\left(\frac{\partial \bs{C}^n_e}{\partial \bs{\xi}^{n-1}_e}\right)^T \bs{\phi}^n_e$ \Comment{compute local history vector}
\State Unseed $\bs{\xi}^{n-1}_e$, seed $\bs{\beta}$, and evaluate $\bs{C}^n_e$ to obtain $\frac{\partial \bs{C}^n_e}{\partial \bs{\beta}}$ \Comment{obtain derivatives with AD}
\State Evaluate $\bs{R}^n_e$ to obtain $\frac{\partial \bs{R}^n_e}{\partial \bs{\beta}}$ \Comment{obtain derivatives with AD}
\State $\objgrad \leftarrow \objgrad +(\bs{\eta}^n_e)^T \frac{\pt \bs{R}^n_e}{\pt \bs{\beta}} +  (\bs{\phi}^n_e)^T \frac{\pt \bs{C}^n_e}{\pt \bs{\beta}}$ \Comment{accumulate gradient}
\State Scatter $\tilde{f}^{n-1}_e$ and $g^{n-1}_e$ \Comment{scatter element contributions to global data}
\EndFor
\EndFor
\end{algorithmic}
\end{spacing}
\label{alg:adjoint_gradient}
\end{algorithm}

Ultimately, the choice to use the forward sensitivities or adjoint method for the gradient will depend on the dimension of the parameter space and memory limitations. Forward sensitivities may be a clear winner when the forward model is high-resolution (e.g. millions of elements) and there are hundreds-thousands of load steps, such that storage of the entire forward solution is prohibitive. The dimension of parameter space in calibration problems studied in the literature is typically small ($<$ 20 parameters), so forward sensitivities may be competitive with the adjoint approach. In topology optimization the dimension of the parameter space scales with that of the state space so forward sensitivities are not viable. 

\section{Results}
\label{sec:results}

In this section, the proposed AD techniques are applied to three different example
problems. The first example provides verification of the AD-computed
gradients and also highlights the efficiency of adjoint-based and forward
sensitivity-based inverse approaches compared to FEMU with finite differences. The second
example illustrates our ability to perform elastoplastic constitutive model calibration for
large-scale problems for which FEMU would be too cost prohibitive
for practical engineering applications. The final example demonstrates the
ease of modification of the AD approach to consider different classes of
constitutive models. In particular, we consider an extension to viscoplasticity.

The examples have common properties, which we outline now. First, the units
have been non-dimension-\\alized to simplify the presentation. Second, the
BCs for each example are of a similar nature. In particular,
each test specimen is pulled in the $y$-direction by prescribing
a load-step dependent traction BC of the form
$\bs{h} = [0, h_y(t), 0]$ on the geometric face with maximal $y$-coordinate
and by prescribing a homogeneous displacement BC $\bs{g} = [0, 0, 0]$
on the geometric face with minimal $y$-coordinate.
All remaining geometric faces are traction-free. Additionally,
the objective function $\obj$ defined by the minimization problem
$\eqref{eq:opt_problem}$ is utilized in each example. Here the surface
$\Gamma_{\text{DIC}}$ in $\obj$ is defined as the geometric face with
maximal $z$-coordinate. Convergence history plots and a table of total objective function and gradient evaluations for each example are provided in \ref{sec:cvg_info}. 

Finally, in each example we examine how the presence of noise degrades
the solution of the inverse problem. We create synthetic \emph{noisy} measurements 
by adding a random draw from a normal distribution with mean zero and unit
variance times a problem-dependent scaling factor $\epsilon_{\text{noise}}$
to each component of the
\emph{noiseless} data at each node, i.e.\ at a given node $d_i = u_i^{\text{true}} + n \epsilon_{\text{noise}}$ with $n \sim \mathcal{N}(0,1)$. The noiseless synthetic data is given by the
forward problem displacements when `exact' material parameters $\bs{\beta}$
are used. We note that the procedure
for generating noisy synthetic data results in a noise level that is the same
in each load step, such that the signal-to-noise (SNR) ratio increases over time
(because the displacement magnitude increases over each load step), which is in line with expectations for experimental DIC measurements.

Two or three noise levels are considered in each problem. The value $\epsilon^*_{\text{noise}}$ is meant to be close to
noise encountered in actual DIC measurements, while the $\epsilon_{\text{noise}}$ value(s) are significantly (i.e. one to two orders of magnitude) larger than $\epsilon^*_{\text{noise}}$ and are meant to provide a sense of how the accuracy of the calibration degrades with decreasing SNR. To estimate $\epsilon^*_{\text{noise}}$, we assume typical parameters for experimental DIC measurements, such that the $y$ extent of the undeformed sample $L_y$ fills 80\% of the field-of-view, the standard deviation of the noise is $0.05$ pixels, and the camera resolution is 5 MP ($2048 \times 2048$). This leads to a value of  $\epsilon^*_{\text{noise}}$ in non-dimensional physical units computed by

\begin{gather}
\epsilon^*_{\text{noise}} = \frac{0.05 L_y}{0.8 \times 2048}.
\end{gather}

\subsection{A Plate with a Cylindrical Hole}

\begin{figure}[ht!]
\centering
\begin{subfigure}{0.25\textwidth}
\vspace{20mm}
\centering
\includegraphics[width=0.95\textwidth]{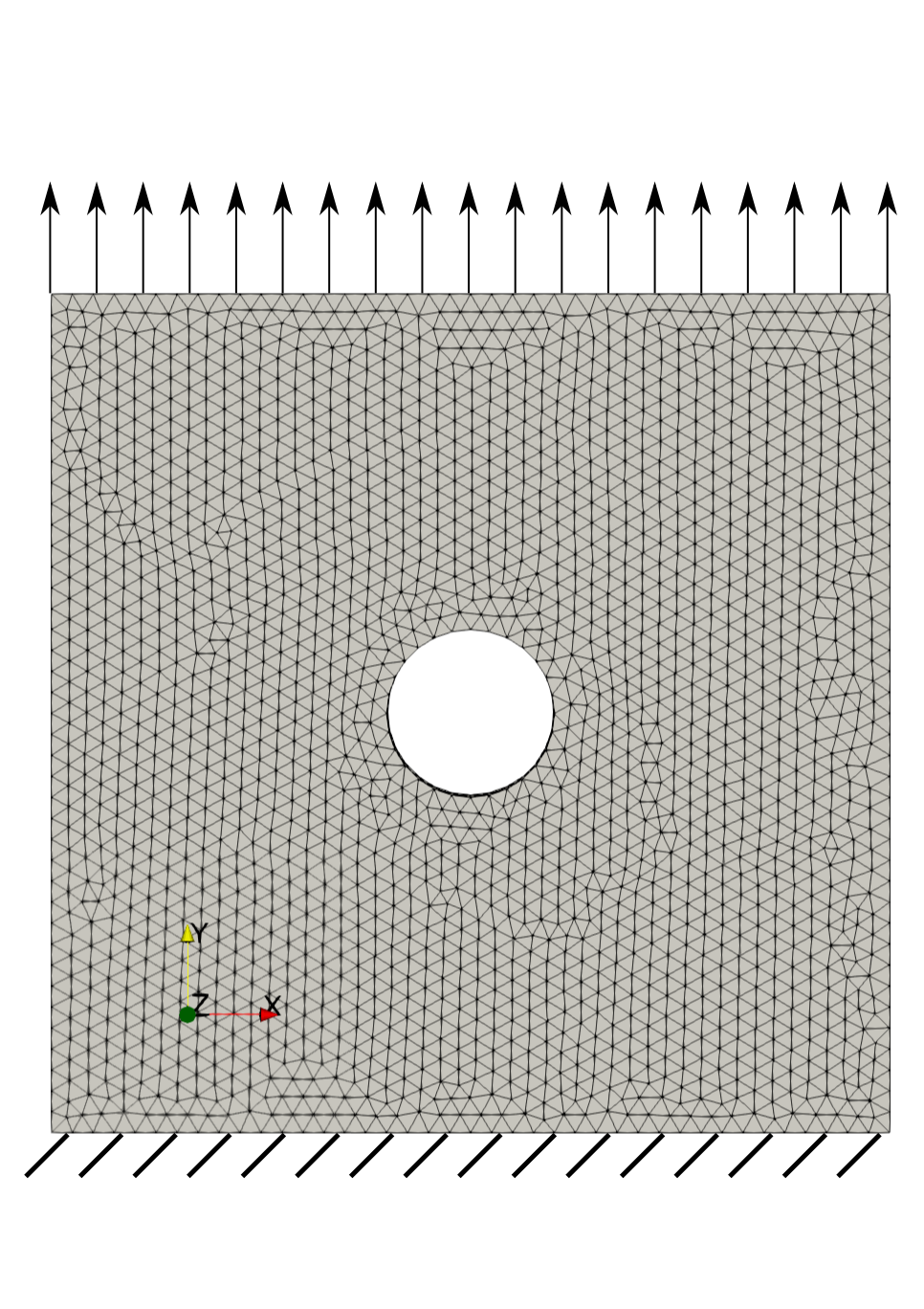}
\vspace{-6mm}
\caption{The geometry, mesh, and boundary conditions $\bs{u} = 0$ on the bottom and
         $h_y = t$ on the top. The DIC surface $\Gamma_{\text{DIC}}$ is the \\ visible plane.}
\label{fig:circle_mesh}
\end{subfigure}%
\hspace{1ex}
\begin{subfigure}{0.3\textwidth}
\centering
\includegraphics[width=0.95\textwidth]{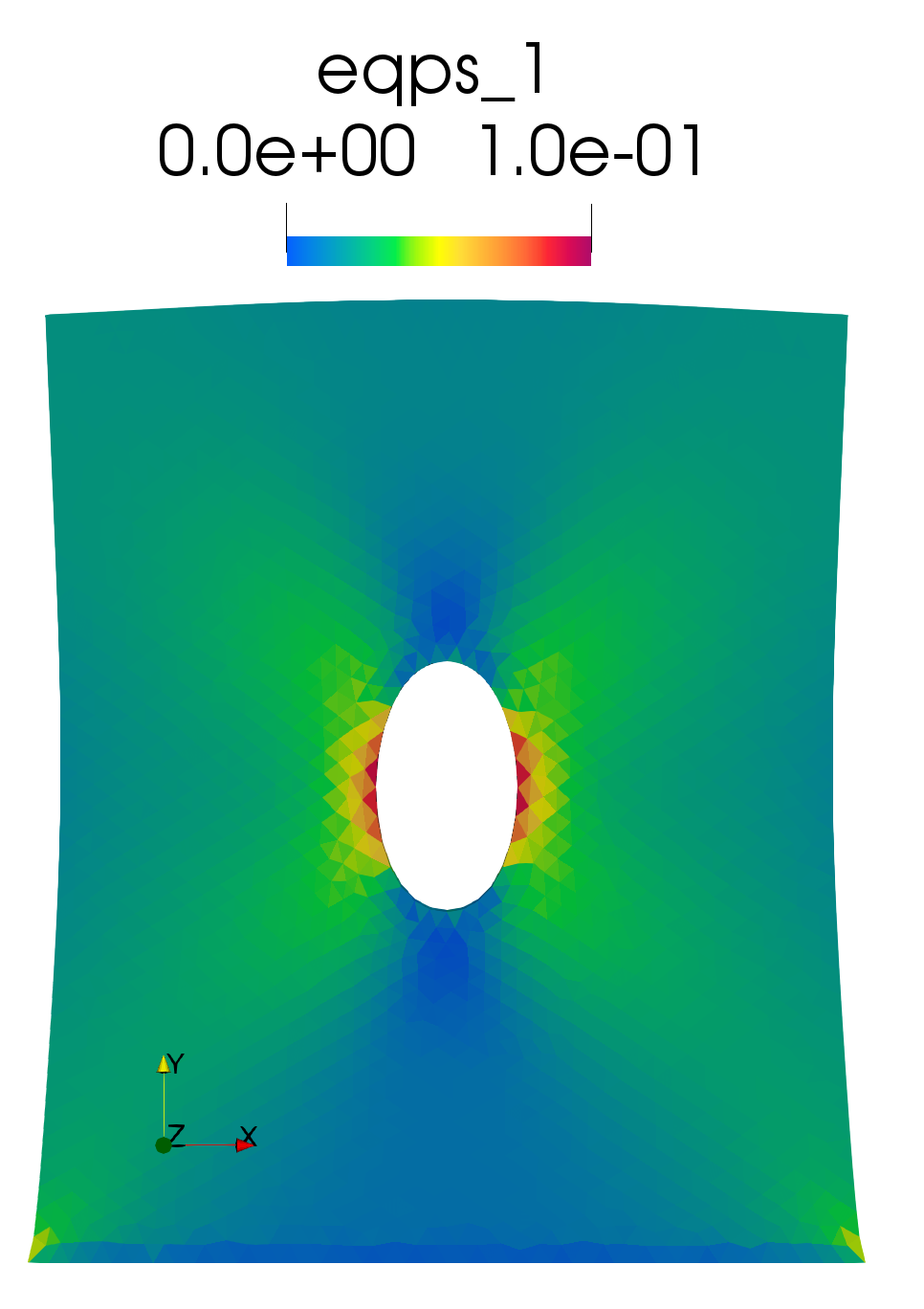}
\caption{The deformed geometry and equivalent plastic strain
at the final load step with displacements exaggerated by a factor of 5.}
\label{fig:circle_eqps_final}
\end{subfigure}\hspace{3mm}%
\begin{subfigure}{0.3\textwidth}
\centering
\includegraphics[width=0.95\textwidth]{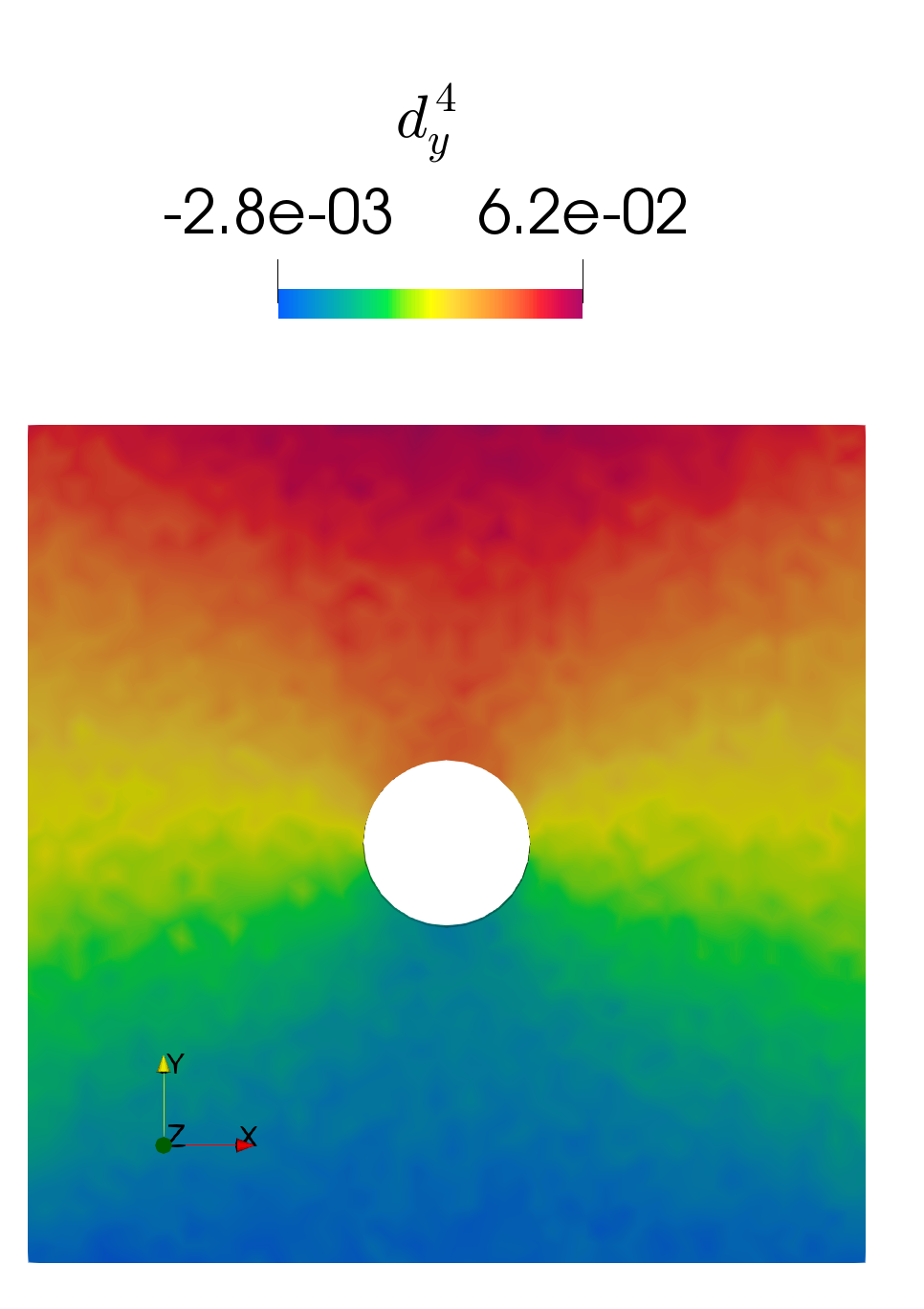}
\caption{The $y$-component of the noisy synthetic displacement
data at the final load step.}
\label{fig:circle_noisy_data}
\end{subfigure}
\caption{Figures for the plate with a cylindrical hole example.}
\end{figure}

In this section, we provide several results that demonstrate the
correctness of the objective function gradients obtained using forward and adjoint sensitivities.
We consider a thin plate
$[-1,1] \times [-1,1] \times [0,0.05]$ with a
cylindrical hole of radius $r = 0.2$ centered at
the origin.
The exact material model parameters are chosen to be
\begin{gather}
\bs{\beta} := [E, \nu, Y, K, S, D] = [1000, 0.25, 2, 100, 0, 0]
\label{eq:circle_exact_params}
\end{gather}
such that the governing constitutive response
\eqref{eq:yield_surface} reduces to one of isotropic
linear hardening. The final two parameters
$S$ and $D$ have been made ``inactive'' for the current results,
meaning they are not considered as design variables.
The domain is discretized using roughly thirteen thousand
tetrahedral elements as shown in Figure \ref{fig:circle_mesh}.
The traction boundary condition is defined by $h_y = t$, and
the forward problem is run for $4$ load steps. Figure
\ref{fig:circle_eqps_final} illustrates an exaggerated version
of the deformed geometry at the final load step.

As an initial step, we consider noiseless synthetic data and
check the validity of the gradients obtained using forward
(Algorithm \ref{alg:forward_sens_gradient})
and adjoint (Algorithm \ref{alg:adjoint_gradient})
sensitivities via a comparison to the finite difference
gradient (Algorithm \ref{alg:fd_gradient}).
Concretely, this occurs by choosing a direction vector $\bs{D}$
and computing an error as the absolute value of the difference
between the automatic differentiation gradient in the direction of
this vector and the two-point centered finite difference gradient
in the direction of this vector. Presently, we choose each
component of this direction vector to be $D_i = 0.1$.
As an example, the error for the gradient obtained via
adjoint sensitivities is computed as:

\begin{gather}
E_{\text{FD\_check}} = \left|
\left( \left[ \objgrad \right]_{\text{adjoint}}
\cdot \bs{D} \right) - 
\left( \left[ \objgrad \right]_{\text{FD}}
\cdot \bs{D} \right)
\right|
\end{gather}

\begin{figure}[ht!]
\centering
\includegraphics[width=0.75\textwidth]{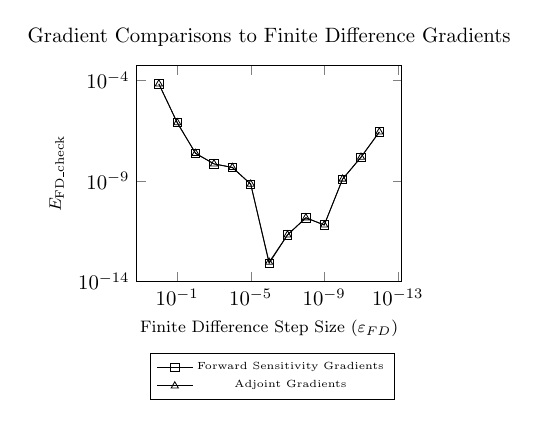}
\caption{Finite difference gradient check for the plate with a
cylindrical hole example.}
\label{fig:circle_vcycle}
\end{figure}

This error is computed using a variety of finite difference step sizes:
$\varepsilon_{\text{FD}} = 1,10^{-1},10^{-2}, \dots, 10^{-12}$ and the
results are plotted in Figure \ref{fig:circle_vcycle}. As is
common with finite difference methods, we see the error in the
finite difference approximation to the gradient decrease as
the step size decreases until an inflection point 
at around $\varepsilon_{\text{FD}} = 10^{-6}$, where
round-off error begins to grow and the finite difference approximation
loses accuracy. This is a known phenomena for computational
finite difference methods, and its presence is a strong indication
that the gradients computed by automatic differentiation are
exact.
Additionally, we remark that the errors
for the gradient obtained with forward and
adjoint sensitivities are numerically equivalent, indicating
that the gradients computed with automatic differentiation are
identical. This lends further credence to the notion that the
correct gradients are computed when using the developed
AD approaches.

As a second verification test, we set the parameters at an
initial guess $\bs{\beta}_0 = [1020, 0.28, 2.3, 110.0, 0, 0]$
that is slightly different than the exact material parameters.
We then use an L-BFGS-B algorithm
to determine the material parameters that solve the optimization problem
\eqref{eq:opt_problem} using gradients obtained with forward finite
differences, forward sensitivities, and adjoint sensitivities.
Here, the lower and upper bounds have been chosen as
$\bs{\beta}_{\text{lo}} = [900, 0.2, 0, 90, 0, 0]$,
and $\bs{\beta}_{\text{hi}} = [1200, 0.4, 10, 150, 0, 0]$, respectively.
For all three approaches, the L-BFGS-B algorithm terminates
with zero percent relative error to at least eight significant digits
for each design parameter. This provides further support of
the correctness of the gradients computed with AD.

\begin{table}[ht!]
\centering
\begin{tabular}{ | c | c | c | c | c | c | }
\hline
  & $E$ & $\nu$ & $Y$ & $K$ & Time \\ \hline
Truth   & 1000 & 0.25 & 2 & 100 & n/a \\ \hline
Initial & 1020 & 0.28 & 2.3 & 110 & n/a \\ \hline
\multicolumn{6}{|c|}{$\epsilon^*_{\text{noise}} =  \num{6.1e-5}$} \\ \hline
FEMU  & 1000.81 (0.081\%) & 0.2493 (0.277\%) &  1.9997 (0.016\%) & 99.9985 (0.002\%) &  103m23s \\ \hline
FS  & 1000.81 (0.081\%) & 0.2493 (0.277\%) &  1.9997 (0.016\%) & 99.9985 (0.002\%) &  21m25s \\ \hline
Adjoint  & 1000.81 (0.081\%) & 0.2493 (0.277\%) &  1.9997 (0.016\%) & 99.9985 (0.002\%) & 14m52s \\ \hline
\multicolumn{6}{|c|}{$\epsilon_{\text{noise}} = \num{1e-3}$} \\ \hline
FEMU    & 1009.22 (0.922\%) & 0.2491 (0.376\%) & 1.9952 (0.234\%) & 100.006 (0.006\%) & 89m38s \\ \hline
FS   & 1009.22 (0.922\%) & 0.2491 (0.376\%) & 1.9952 (0.234\%) & 100.006 (0.006\%) &    18m43s \\ \hline
Adjoint & 1009.22 (0.922\%) & 0.2491 (0.376\%) & 1.9952 (0.234\%) & 100.006 (0.006\%) &13m12s \\ \hline
\end{tabular}
\caption{Inverse problem solutions for the plate with a cylindrical hole
problem with noisy synthetic data. Errors relative to the true solutions
are reported next to the parameter values.}
\label{table:circle_results}
\end{table}

As a final demonstration, we consider two sets of noisy synthetic data with $\epsilon^*_{\text{noise}} = \num{6.1e-5}$
and $\epsilon_{\text{noise}} = \num{1e-3}$. The $y$-component of the noisier displacement dataset 
is shown in the right panel of Figure \ref{fig:circle_noisy_data}. The optimization
problem \eqref{eq:opt_problem} (with initial guess $\bs{\beta}_0$ and bounds
$\bs{\beta}_{\text{lo}}$ and $\bs{\beta}_{\text{hi}}$) is then solved
using finite difference (FEMU), forward sensitivity (FS), and adjoint-based gradients.
Results obtained using the three methods are shown in Table \ref{table:circle_results}.
We observe that to the precision shown in the table each method obtains the same solution,
but the newly developed AD-based approaches take roughly one fifth the time of the
FEMU approach. As mentioned in the previous section, this cost savings comes from
the replacement of full nonlinear solves in the FEMU approach with the auxiliary
linear solves present in the other methods.

Lastly, we also investigated the sensitivity of the solution of the inverse problem to the initial guess
and amount of noise in this example. Both noise-corrupted inverse problems were run using initial guesses
generated from the first ten elements of the 2,3,5,7 Halton sequence transformed by the optimization bounds (a ``well-spaced'' set of samples). We found that for each noise level six of these initial guesses resulted in the same solutions given in Table \ref{table:circle_results}. The other four initial guesses had a $Y >= 6$, and consequently produced forward solutions with comparatively smaller amounts of plastic deformation in the material. In some cases there was no plastic deformation until the last one or two load steps, and while the magnitude of the equivalent plastic strain was non-zero it was still diminutive and only present in a small percentage of elements. The reduction in the value of the objective function for these problems was extremely slow compared to that witnessed for the other initial guesses. In light of these observations, we recommend choosing a value of $Y$ that results in an appreciable amount of plastic strain in the material when the goal is to calibrate an elastoplastic constitutive model. These results also suggest that the minimum is unique over the range of optimization bounds considered in this example. In practice these bounds should be determined by a combination of expert knowledge (e.g.\ appropriate ranges for steel variants) and trial-and-error based on optimizer performance, as it is conceivable that for some problem configurations (i.e.\ specimen design, loading, constitutive models, calibration parameters, data quality) the optimization problem will be non-convex.

\subsection{Large Problem}

\begin{figure}[ht!]
\centering
\begin{subfigure}{0.3\textwidth}
\centering
\includegraphics[width=0.95\textwidth]{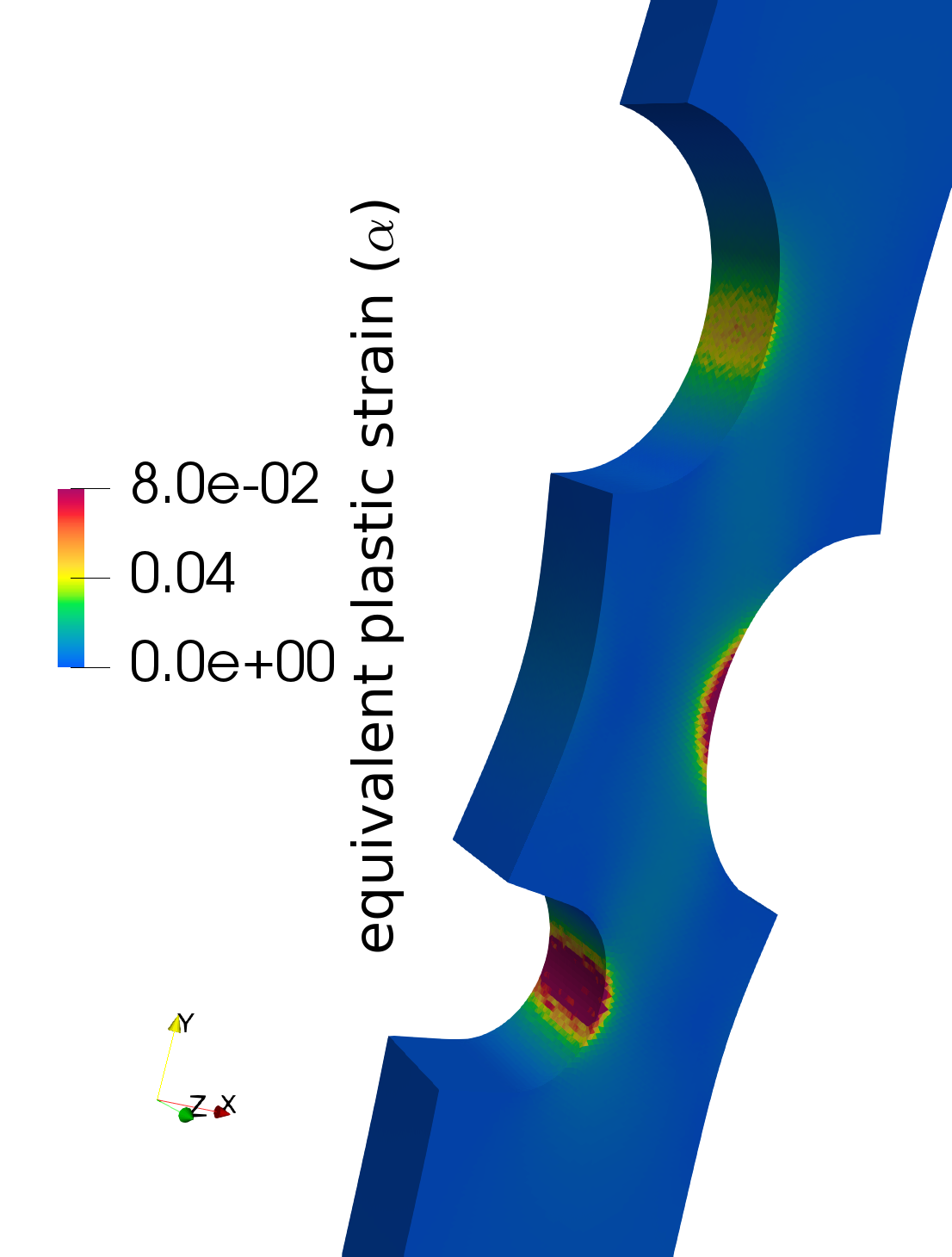}
\caption{The deformed geometry and the equivalent plastic strain at the
final (tenth) load step.}
\label{fig:xprize_eqps}
\end{subfigure}\hspace{2mm}%
\begin{subfigure}{0.3\textwidth}
\centering
\includegraphics[width=0.95\textwidth]{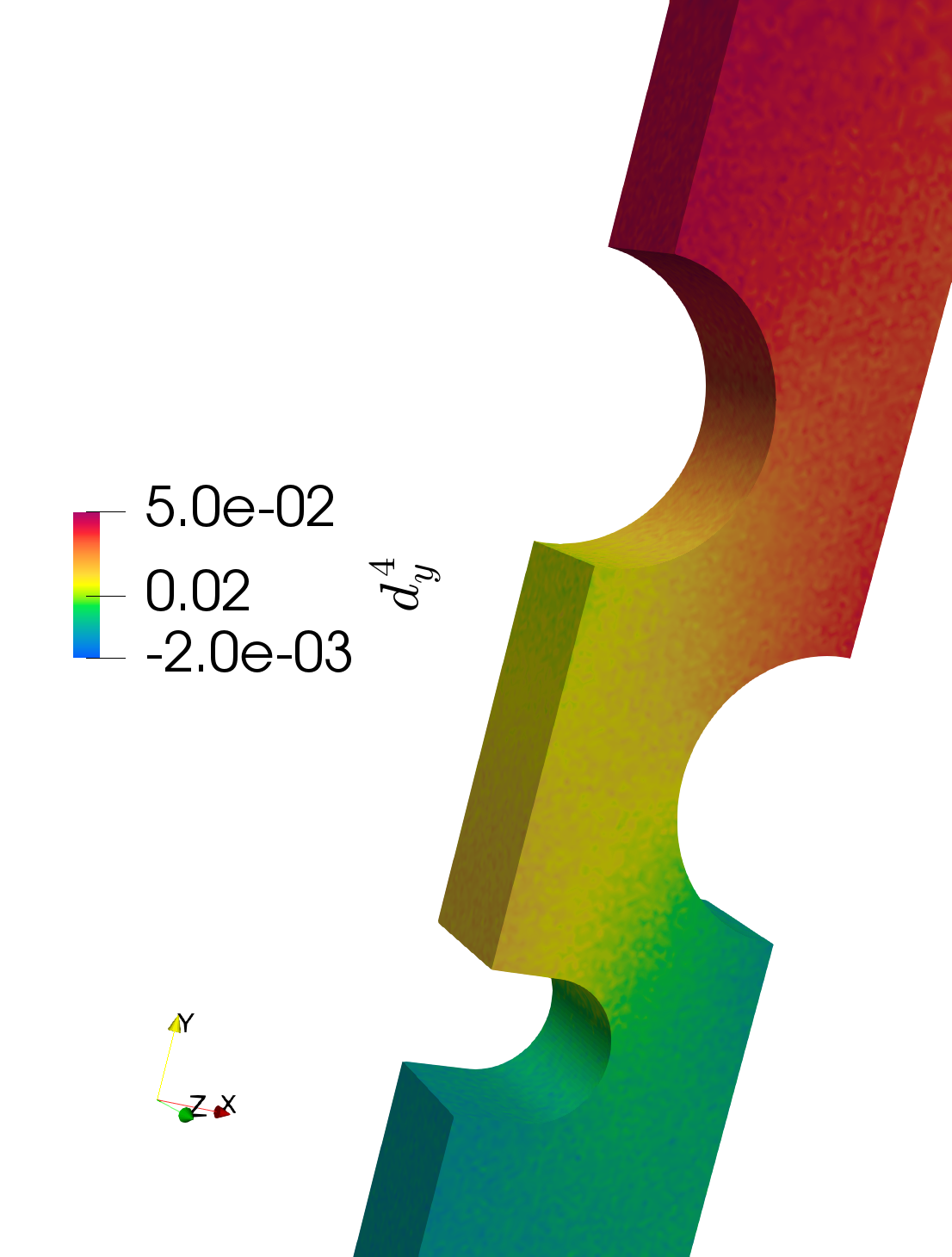}
\caption{The $y$-component of the noisy synthetic data computed with
$\epsilon_{\text{noise}} = 0.001$ at the fourth load step.}
\label{fig:xprize_lower_noise}
\end{subfigure}\hspace{2mm}%
\begin{subfigure}{0.3\textwidth}
\centering
\includegraphics[width=0.95\textwidth]{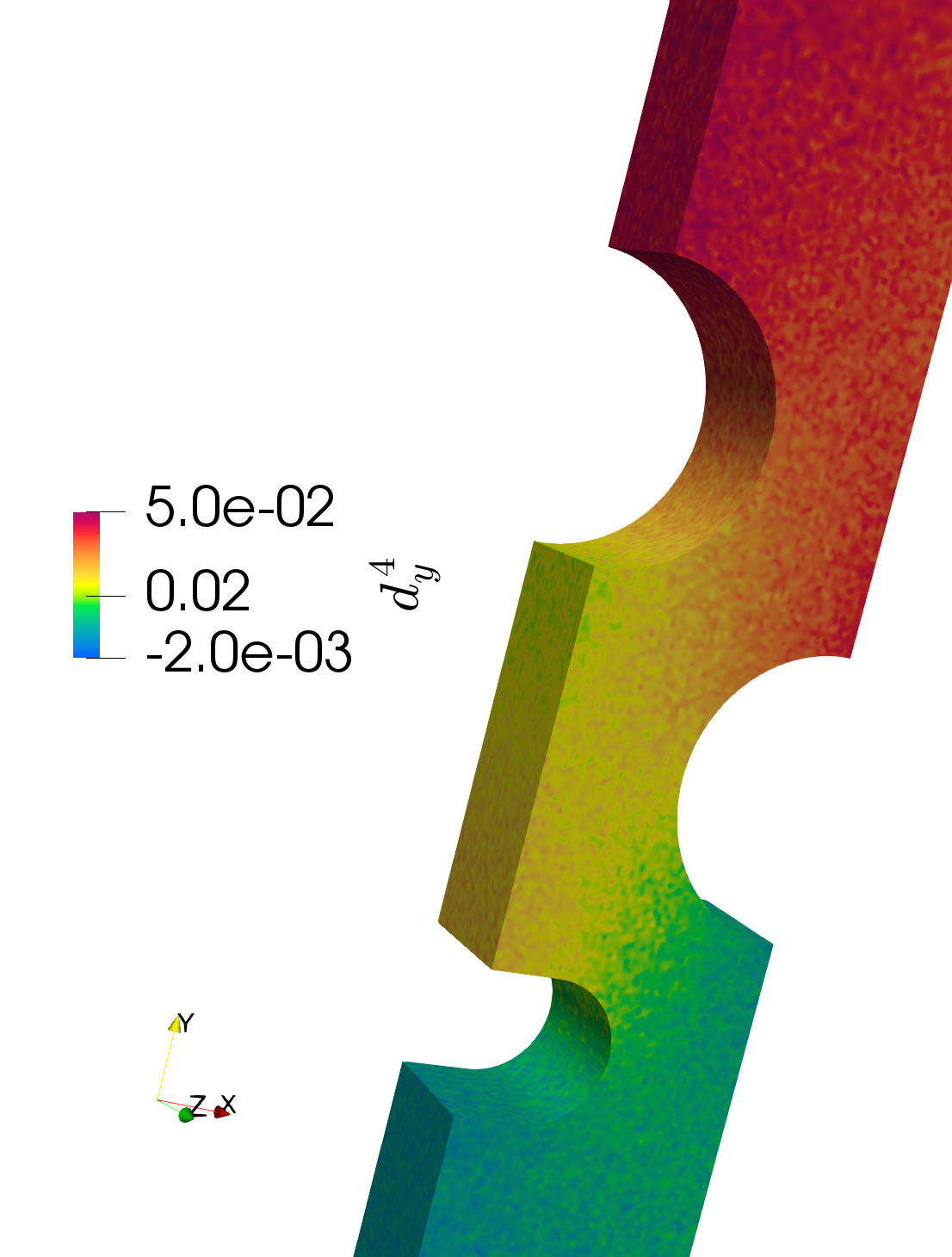}
\caption{The $y$-component of the noisy synthetic data computed with
$\epsilon_{\text{noise}} = 0.002$ at the fourth load step.}
\label{fig:xprize_higher_noise}
\end{subfigure}
\caption{Figures for the large example problem.} 
\end{figure}

In this section we demonstrate that the proposed AD-based adjoint inversion
approach can be used to perform material model calibration for large-scale
simulations that would be otherwise intractable using finite difference-based FEMU. In particular,
we consider a rectangular prism
specimen $[0,0.25] \times [-1.625,1.625] \times [0, 0.124]$
with three notches of varying sizes near the middle of the specimen (described in \cite{ostien201610})
that introduce heterogeneity into the response. Figure \ref{fig:xprize_lower_noise}
shows a picture of the undeformed sample. The geometry is discretized with
over one million tetrahedral elements.

The forward problem is defined by the traction boundary condition $h_y = 2t$
and was run for a total of ten load steps, with exact material parameters
defined as $\bs{\beta} = [E, \nu, Y, S, D] = [1000, 0.32, 6, 120, 100]$.
Here the linear hardening parameter $K$ has been set to zero and non-zero
values for $S$ and $D$ have been specified to induce Voce hardening in
the specimen. The equivalent plastic strain $\alpha$
on the deformed geometry is shown in Figure \ref{fig:xprize_eqps}.
We consider three sets of noisy synthetic data, created using $\epsilon^*_{\text{noise}} = \num{1e-4}$, $\epsilon_{\text{noise}} = \num{1e-3}$, and $\epsilon_{\text{noise}} = \num{2e-3}$.
For the two noisier sets, the $y$-component of the noisy displacement data at
the fourth load step is shown in Figures
\ref{fig:xprize_lower_noise} and \ref{fig:xprize_higher_noise}.
In subsequent load steps, it is more difficult to see the noise in the data.

The optimization problem \eqref{eq:opt_problem} was solved using
adjoint-based gradients for all sets of noisy synthetic data. The initial
parameters for the inverse problem were chosen to be
$\bs{\beta}_0 = [1200, 0.36, 3.5, 105, 80]$, with lower and upper bounds
specified as
$\bs{\beta}_{\text{lo}} = [800, 0.25, 3, 100, 60]$ and
$\bs{\beta}_{\text{hi}} = [1600, 0.4, 10, 150, 200]$, respectively.
For these model calibration runs, the parameter $K$ was been made ``inactive''.
The inverse problems were solved in parallel using 64 MPI ranks, and took roughly 9 or 12 hours to run using the
adjoint and forward sensitivities approaches, respectively.

\begin{table}[ht!]
\centering
\begin{tabular}{ | c | c | c | c | c | c | }
\hline
              & $E$  & $\nu$  & $Y$   & $S$ & $D$  \\ \hline
Truth         & 1000 & 0.32   & 6     & 120 & 100  \\ \hline
Initial       & 1200 & 0.36   & 3.5   & 105 & 80   \\ \hline
$\epsilon^*_{\text{noise}}$ = $\num{1e-4}$ & 999.9 (0.006\%) & 0.32 (0.09\%) & 5.996 (0.06\%) & 119.95 (0.042\%) & 100.18 (0.18\%) \\ \hline
$\epsilon_{\text{noise}}$ = $\num{1e-3}$ & 998.8 (0.13\%) & 0.33 (1.74\%) &  6.17 (2.75\%) & 119.94 (0.046\%) & 100.5 (0.50\%) \\ \hline
$\epsilon_{\text{noise}}$ = $\num{2e-3}$ & 1011.5 (1.15\%) & 0.31 (4.13\%) & 4.70 (21.7\%) & 121.18 (0.98\%) & 92.94 (7.06\%) \\ \hline
\end{tabular}
\caption{Inverse problem solutions for the large-scale example 
problem with noisy synthetic data. Errors relative to the true solutions
are reported next to the parameter values.}
\label{table:xprize_results}
\end{table}

The results of the inverse problems are given in Table \ref{table:xprize_results}.
As expected, the accuracy of the solution decreases as the amount of random noise
in the data is increased. However, we note that each parameter is affected differently.
For example, the saturation rate $D$ appears to be more sensitive to the presence of
noise than the saturation modulus $S$, which suggests that the overall sensitivity of
these parameters given the ``experimental'' setup is not equal.
Although we do not explore this line of investigation in this work,
global sensitivity analyses could be useful in designing experiments that
exhibit high (and possibly more equal) sensitivity to all of the parameters of interest.

Finally, we remark that our demonstration of inversion with a mesh consisting of over one million
elements is larger than we have encountered in the full-field model calibration literature.
This is most likely due to the costs incurred by finite difference approximations.
Ten load steps, however, is a rather small amount of DIC data when compared to the literature.
Given that the focus of this paper is a demonstration of feasibility, we leave further exploration
of the HPC aspects to future work. This work could include tackling the challenges of using
many load steps, along with other technical hurdles that we elaborate on in
Section \ref{sec:discussion}.

\subsection{Hypoelastic Model with Anisotropic Yield}

As a final example, we showcase how our AD-based framework makes the implementation of finite deformation elastoplastic FE model with the Hill anisotropic yield function and elastic behavior described by a hypoelastic model below yield nearly trivial, as all that is required is an appropriate specification of the global and local residuals. In this section we present these residuals and then utilize them in a FE model that emulates a mechanical characterization experiment for a bi-axial test specimen with non-proportional loading where the goal is to estimate the in-plane anisotropy coefficients given synthetic DIC measurements with noise levels commensurate with those found in real experimental data.

Our use of a hypoelastic model in this example is motivated by several factors. First, some well-known disadvantages of these models (as compared to hyperelastic models) are their non-conservative nature and required use of an incrementally-objective integration algorithm for the rate of stress. The energy error introduced by the former, however, is negligible when the elastic strains are small compared to the plastic strains \cite{belytschko2013nonlinear}, as is the case in many problems of practical importance (e.g.\ sheet metal forming). Second, the proper treatment of anisotropic plasticity (i.e.\ anisotropic yield and/or distortional hardening) is arguably more challenging when described by a hyperelastic model \cite{vladimirov2010anisotropic}, whereas with hypoelastic formulations it is straightforward to develop finite deformation versions of such models from small strain descriptions. Lastly, the presentation of a hypoelastic model in this section serves to highlight both the applicability of our AD-based approach to both classes of model formulations (encompassing all of those used in practice) while demonstrating relative ease with which it can be applied to a specific model of interest.

The global residual $\bs{R}$ is similar to \eqref{eq:eq_res} but due to our use of a hypoelastic model we can
no longer derive expressions for the Kirchhoff stress and pressure from a strain-energy density function.
We instead introduce a split in the hydrostatic and deviatoric parts of the Cauchy stress 

\begin{gather}
\bs{\sigma} = \text{dev}(\bs{\sigma}) + \frac{\text{tr}(\bs{\sigma})}{3} \bs{I}
            = \text{dev}(\bs{\sigma}) - p \bs{I} \label{eq:cauchy_pressure_split}
\end{gather}

\noindent to allow for the 
equal-order interpolation of the displacement and pressure fields via stabilization as was done in the hyperelastic-plastic formulation presented previously.

We change the local state variables vector to include the \emph{unrotated} Cauchy stress
$\bs{T}$ and equivalent plastic strain $\alpha$ such that $\bs{\xi} := \{\bs{T}, \alpha\}$. Due
to the symmetry of $\bs{T}$, $\bs{\xi}$ has seven independent components. The Cauchy stress
$\bs{\sigma}$ and $\bs{T}$ are related through the polar decomposition of the deformation
gradient

\begin{gather}
\begin{aligned}
\bs{F} &= \bs{R} \bs{U}, \\
\bs{T} &= \bs{R}^T \bs{\sigma} \bs{R}. \label{eq:unrot_Cauchy}
\end{aligned}
\end{gather}

The stabilized global residual $\bs{R}$ is

\begin{gather}
\begin{split}
&\int_{\B} (J \ (\text{dev}\left[\bs{R} \bs{T} \bs{R}^T\right] - p \bs{I}) \ \bs{F}^{-T}) : \nabla \bs{w} \, \text{d} V -
\int_{\B} \left( p + \frac{\text{tr} \left[ \bs{R} \bs{T} \bs{R}^T \right]}{3} \right) q \, \text{d} V - \\
&\sum_{e=1}^{n_{el}} \int_{\B_e} \tau_e (J \bs{F}^{-1} \bs{F}^{-T}) : (\nabla p \otimes \nabla q) \, \text{d} V -
\int_{\Gamma_h} \bs{h} \cdot \bs{w} \,\, \text{d} A = 0,
\end{split} \label{eq:eq_res_hypo}
\end{gather}

\noindent where all of the quantities are discretized at the current time step (superscript $n$ omitted for brevity),
and the finite element spaces are the same as those stated previously in \eqref{eq:fe_spaces}.

Following \cite{lame_manual} and the references therein, we utilize the Green-McInnis stress rate for objective
integration of the rate of the Cauchy stress through a \emph{corotational formulation}, which involves the use of the polar decomposition of
the deformation gradient at the current time step. The relevant kinematics expressed in discrete form using
a backward Euler discretization are

\begin{gather}
\begin{aligned}
\bs{F}^n &= \bs{R}^n \bs{U}^n, \\
\bs{L}^n &= \left( \bs{F}^n - \bs{F}^{n-1} \right) \left( \bs{F}^n \right)^{-1}, \\
\bs{D}^n &= \frac{1}{2} \left( \bs{L}^n + \left( \bs{L}^n \right)^T \right), \\
\bs{d}^n &= \left(\bs{R}^n\right)^T \bs{D}^n \bs{R}^n, \label{eq:hypoelastic_kinematics}
\end{aligned}
\end{gather}

\noindent where $\bs{L}$ is the velocity gradient, $\bs{D}$ is the rate of deformation tensor, and $\bs{d}$ is the \emph{unrotated}
rate of deformation tensor.

Hypoelastic constitutive models assume an additive decomposition of $\bs{d}$ into elastic $\bs{d}^e$ and plastic $\bs{d}^p$ components. The backwards Euler temporal discretization of the constitutive equation for an isotropic hypoelastic material expressed in rate form in the unrotated configuration is

\begin{gather}
\begin{aligned}
\bs{d}^n &=  \left(\bs{d}^e \right)^n  + \left(\bs{d}^p \right)^n, \\
\bs{T}^n &= \bs{T}^{n-1} + \Delta t \left( \lambda \text{tr}\left[ \left( \bs{d}^e \right)^n \right] \bs{I} + 2 \mu  \left( \bs{d}^e \right)^n \right). \label{eq:hypoelastic_rate_ce}
\end{aligned}
\end{gather}

The Hill \cite{hill1948theory} effective stress function $\phi$  contains six coefficients $F, G, H, L, M,$ and $N$

\begin{gather}
\phi(T_{ij}) = \left( F \left( T_{22} - T_{33} \right)^2
                     + G \left( T_{33} - T_{11} \right)^2
                     + H \left( T_{11} - T_{22} \right)^2
                     + 2 L \left(T_{23}\right)^2 + 2 M \left(T_{13}\right)^2 + 2 N \left(T_{12}\right)^2 \right)^{\frac{1}{2}}, \label{eq:Hill_yield}
\end{gather}

\noindent that are defined in terms of a reference yield stress $Y$ and 6 independent axial $\sigma^y_{ii}$ and shear $\tau^y_{ij}$ yield stresses such that

\begin{gather}
\begin{aligned}
F &= \frac{Y^2}{2} \left( \left(\sigma^y_{22} \right)^{-2} + \left(\sigma^y_{33} \right)^{-2} - \left(\sigma^y_{11} \right)^{-2} \right),
\quad  L = \frac{Y^2}{2} \left(\tau^y_{23} \right)^{-2}, \\
G &= \frac{Y^2}{2} \left( \left(\sigma^y_{33} \right)^{-2} + \left(\sigma^y_{11} \right)^{-2} - \left(\sigma^y_{22} \right)^{-2} \right),
\quad M = \frac{Y^2}{2} \left(\tau^y_{13} \right)^{-2}, \\
H &= \frac{Y^2}{2} \left( \left(\sigma^y_{11} \right)^{-2} + \left(\sigma^y_{22} \right)^{-2} - \left(\sigma^y_{33} \right)^{-2} \right),
\quad N = \frac{Y^2}{2} \left(\tau^y_{12} \right)^{-2}.
\end{aligned}
\end{gather}

\noindent For calibration purposes we express the constitutive model parameters in terms of the six independent quantities $R_{ij}$ defined to be

\begin{gather}
R_{ii} = \frac{\sigma_{ii}^y}{Y} \quad \text{for} \ i = 1,2,3,
\qquad R_{ij} = \sqrt{3} \frac{\tau^y_{ij}}{Y} \quad \text{for} \ i,j = 1, 2, 3, \ \text{and} \ i < j.
\end{gather}

\noindent The Hill yield function \eqref{eq:Hill_yield} is equivalent to the $J_2$ yield function when $R_{ij} = 1 \ \forall i,j$. The discretized flow rule and hardening law may be combined to obtain
\begin{gather}
\left(\bs{d}^p \right)^n = \Delta \gamma^n \frac{\pt \phi^n}{\pt \bs{T}^n} = \frac{1}{\Delta t} \left( \alpha^n - \alpha^{n-1} \right) \frac{\pt \phi^n}{\pt \bs{T}^n}.
\end{gather}

\noindent We note that for this model $\text{tr}\left[ \left(\bs{d}^p \right)^n \right] = 0$ \cite{de2011computational}. 

In this final example we assume isotropic linear hardening. The value of the yield function for the trial elastic step $f^n_{\text{trial}} = \phi^n_{\text{trial}} - Y - K \alpha_{\text{trial}}$ is obtained by assuming the rate of deformation is completely elastic (i.e.\ $\bs{d}^n = \left(\bs{d}^e \right)^n$) in \eqref{eq:hypoelastic_rate_ce} and $\alpha_{\text{trial}} = \alpha^{n-1}$. The discrete constitutive equation evolution equations $\bs{C}^n$ below yield (i.e.\ $f^n_{\text{trial}} < 0$) are

\begin{gather}
\begin{cases}
\begin{aligned}
&\bs{T}^n -\bs{T}^{n-1} - \Delta t \left( \lambda \text{tr}\left[ \bs{d}^n \right] \bs{I} + 2 \mu \bs{d}^n \right) = \bs{0}, \\
&\alpha^n - \alpha^{n-1} = 0.
\end{aligned}
\end{cases}
\end{gather}

\noindent When plastic flow occurs they are

\begin{gather}
\begin{cases}
\begin{aligned}
&\bs{T}^n - \bs{T}^{n-1} - \Delta t \left( \lambda \text{tr}\left[ \bs{d}^n \right] \bs{I} + 2 \mu \left( \bs{d}^n -  \frac{1}{\Delta t} \left( \alpha^n - \alpha^{n-1} \right) \frac{\pt \phi^n}{\pt \bs{T}^n} \right) \right) = \bs{0},  \\
&\phi^n(\bs{T}^n) - Y - K \alpha^n = 0.
\end{aligned}
\end{cases}
\end{gather}

To summarize, the hypoelastic-plastic model with anisotropic yield utilizes global and local state residuals that are distinct from those described previously in this work, and the definition of the local state variables is different.

\begin{figure}[ht!]
\centering
\begin{subfigure}{0.3\textwidth}
\centering
\includegraphics[width=0.95\textwidth]{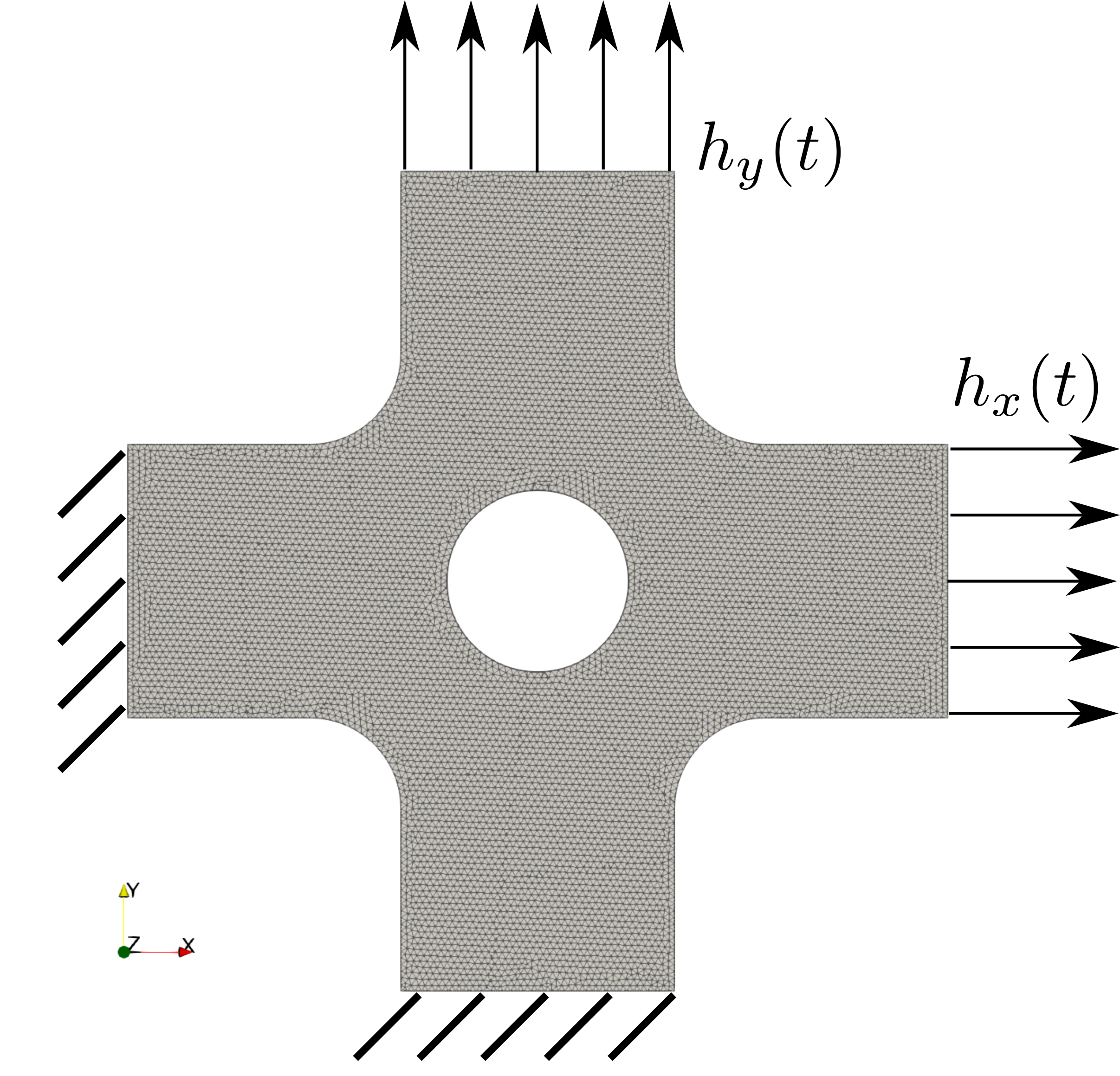}
\caption{The geometry, mesh, and boundary conditions $\bs{u} = \bs{0}$ on the bottom and left edges, $h_y(t)$ on the top edge, and $h_x(t)$ on the right edge. The DIC surface $\Gamma_{\text{DIC}}$ is the visible plane.}
\label{fig:cruciform_mesh}
\end{subfigure}%
\hspace{1ex}
\begin{subfigure}{0.3\textwidth}
\centering
\includegraphics[width=0.95\textwidth]{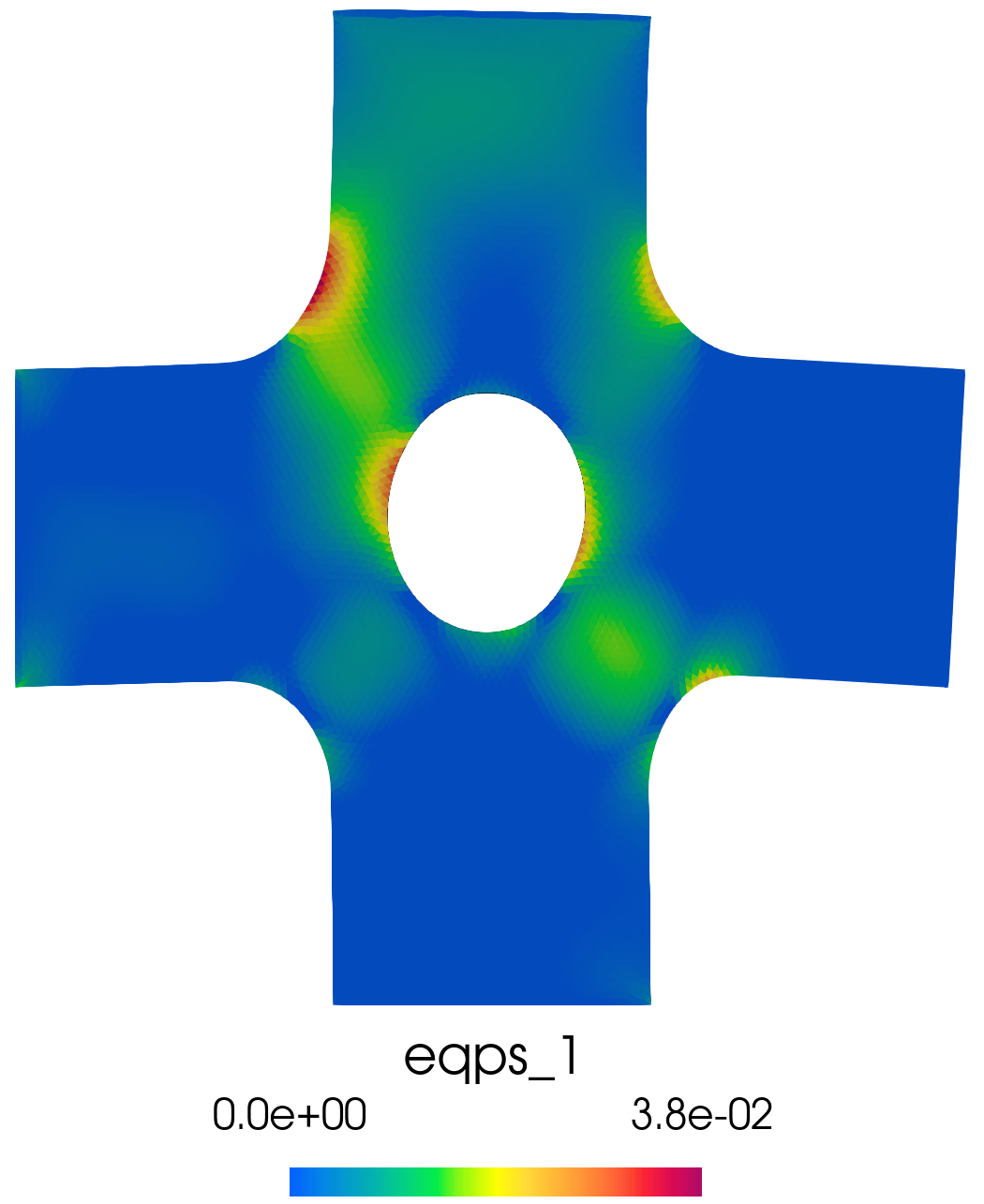}
\vspace{2mm}
\caption{The deformed geometry and equivalent plastic strain
at the 6th load step with displacements exaggerated by a factor of 5.}
\label{fig:cruciform_eqps_mid}
\end{subfigure}\hspace{3mm}%
\begin{subfigure}{0.3\textwidth}
\centering
\includegraphics[width=0.95\textwidth]{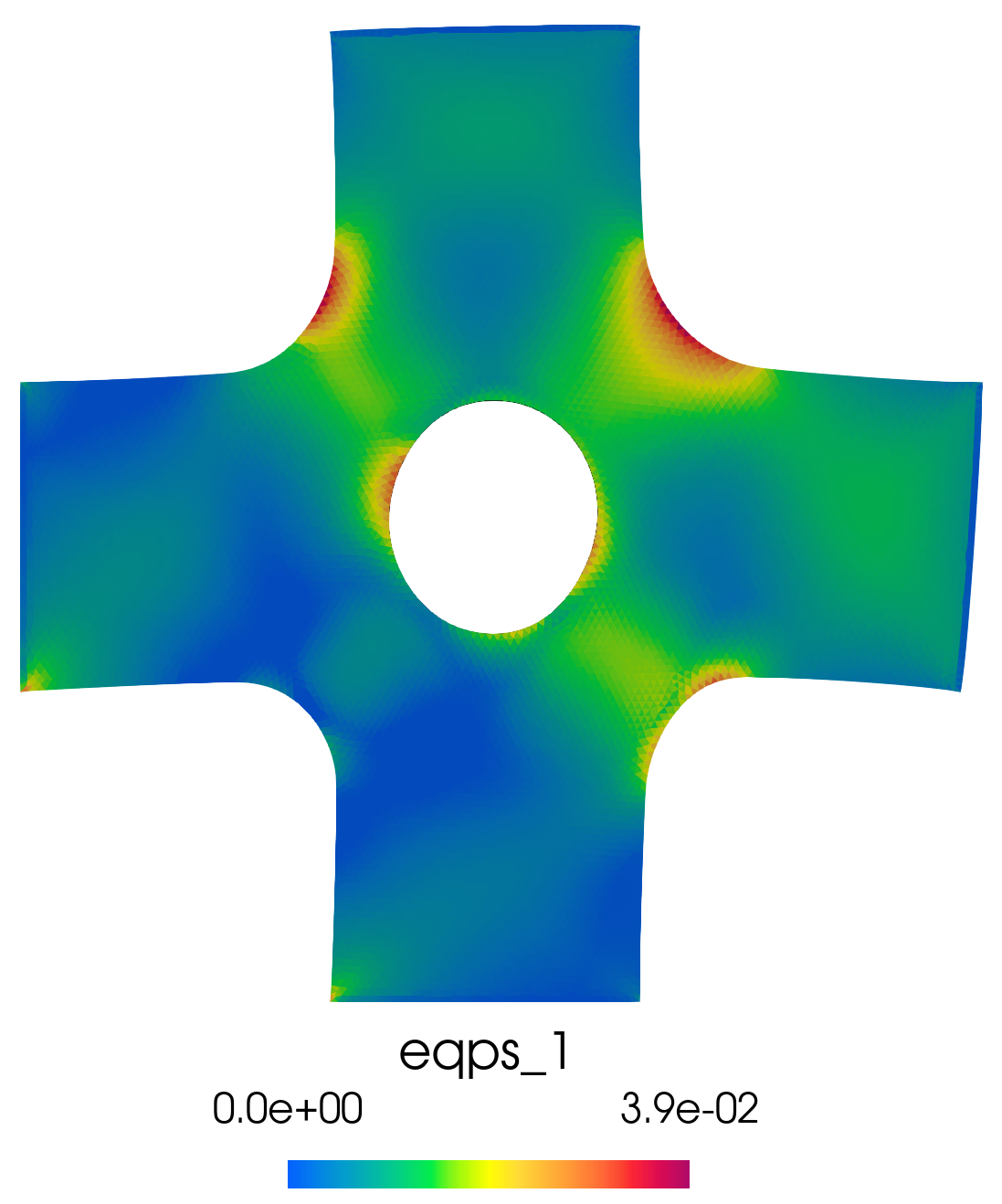}
\vspace{2mm}
\caption{The deformed geometry and equivalent plastic strain
at the 12th load step with displacements exaggerated by a factor of 5.}
\label{fig:cruciform_eqps_final}
\end{subfigure}
\caption{Cruciform specimen geometry, FE model boundary conditions, and equivalent plastic strain under axial (middle) and bi-axial (right) loading.}
\end{figure}

We now present a FE model of a synthetic mechanical characterization experiment that consists of a cruciform test specimen subject to bi-axial loading with the aim of demonstrating the feasibility of our AD-based calibration methods for estimating parameters related to anisotropic yield. We utilize a cruciform specimen geometry that has been adapted from \cite{coppieters2018inverse}. The dimensions of the bounding box that encloses it are $[-4.5, 4.5] \times [-4.5, 4.5] \times [0, 0.12]$ and the radii of the circle in the center and fillets are all equal to 1. The test specimen was deformed non-proportionally over a series of 12 load steps according to the traction boundary conditions

\begin{gather}
\begin{aligned}
&h_x = 0, \ h_y = 0.5 t  &&\text{for} \ t = 1, \ldots, 6, \\
&h_x = 0.5 (t - 6), \ h_y  = 3 \quad &&\text{for} \ t = 7, \ldots, 12.
\end{aligned}
\end{gather}

\noindent that are applied over the surfaces of the test specimen as depicted in Figure \ref{fig:cruciform_mesh}.

In this bi-axial example we set $R_{11} = 1$ so that it is equal to the reference yield stress value $Y$. We also fix the out of plane shear  parameters $R_{23}$ and $R_{13}$ to 1 because the out of plane shear stress in this problem is negligible compared to the in-plane stresses. The constitutive model parameters in this example are 

\begin{gather}
\bs{\beta} = \left[E, \nu, Y, K, R_{11}, R_{22}, R_{33}, R_{23}, R_{13}, R_{12} \right] =  \left[1000, 0.25, 2, 100, 1, 1.05, 0.95, 1, 1, 0.85 \right].
\end{gather}

Calibration results obtaining using finite difference FEMU and forward and adjoint sensitivities are reported in Table \ref{table:cruciform_parameter_results}. All of the calibration problems were executed in parallel using 16 MPI ranks. The running times for the forward and adjoint sensitivities approaches were within minutes of each other for each noise case. They were (on average) 3.8 hours and 7.2 hours for the adjoint and forward sensitivities approaches, respectively. The optimization iteration count and number of objective function and gradient evaluations were the same for these methods, and this finding is consistent with our previous examples. The running times for the FEMU calibration problems, however, exhibited greater variability due to the finite difference approximation of the gradient although their parameter estimates were nearly identical to those obtained using the other two methods. The $\epsilon^*_{\text{noise}} =  \num{2.7e-4}$ and $\epsilon_{\text{noise}} = \num{1e-3}$ FEMU problems took 36 hours and 29 hours, respectively.

\begin{table}[ht!]
\centering
\begin{tabular}{ | c | c | c | c | c | c | c | c | }
\hline
  & $E$ & $\nu$ & $Y$ & $K$ & $R_{22}$ & $R_{33}$ & $R_{12}$ \\ \hline
Truth   & 1000 & 0.25 & 2 & 100 & 0.9 & 1.05 & 0.85 \\ \hline
Initial & 1020 & 0.28 & 2.3 & 110 & 0.95 & 0.95 & 0.95 \\ \hline
\multicolumn{8}{|c|}{$\epsilon^*_{\text{noise}} =  \num{2.7e-4}$} \\ \hline
Parameter Value & 1000.30 & 0.2504 &  1.9996 & 100.053 & 0.89997 & 1.0501 & 0.8501 \\ \hline
Parameter Error (\%) & 0.03 & 0.153 & 0.019 & 0.053 & 0.001 & 0.008 & 0.018 \\ \hline
\multicolumn{8}{|c|}{$\epsilon_{\text{noise}} = \num{1e-3}$} \\ \hline
Parameter Value & 998.92 & 0.2482 & 2.0007 & 99.879 & 0.89976 & 1.0494 & 0.8502 \\ \hline
Parameter Error (\%) & 0.108 & 0.724 & 0.035 & 0.121 & 0.026 & 0.061 & 0.027 \\ \hline
\end{tabular}
\caption{Inverse problem solutions for the cruciform specimen
problem with noisy synthetic data. Solutions obtained through each of the three inverse methods
were the same to the precision shown in the table.}
\label{table:cruciform_parameter_results}
\end{table}

The accuracy of the recovered anisotropic coefficients is encouraging (all $R_{ij}$ are recovered to within 0.1 \% error for both noise cases). These results and those obtained from our previous examples have given us confidence in our numerical formulation and its computational implementation and provided motivation to further develop our approach so that it may be applied to experimental DIC data.

\section{Discussion}
\label{sec:discussion}

In this section we discuss a few aspects of our approach that would need to be extended for calibration with experimental DIC data to be successful. First, as discussed in section \ref{sec:inverse}, the objective function in this work should be augmented with a force-matching term, and weighting functions are needed to properly account for the uncertainty in measurements of displacement and load. Second, the DIC algorithm filters the true displacement signal, and in many cases the data is projected onto the mesh/basis used in the FE model, which can corrupt the objective function in some instances \cite{lava2020validation}. Finally, in our formulation we synchronized load steps and DIC data frames and utilized idealized boundary conditions, and these restrictions should be relaxed. 

Our formulation of the objective function is appropriate when the noise in each displacement measurement is i.i.d.\ Gaussian across all frames. The noise in our example problems conformed to this, but in real DIC measurements the noise exhibits spatial heterogeneity in some instances. In particular, the largest source of this variability occurs in local or subset-based DIC formulations, where the error in the projection of the DIC measurements onto the FE mesh ($\bs{d}$) is considerably higher near the boundaries of specimen. One way to address this issue is to decrease the influence of measurements in these regions through the use of a weighting function in the objective function that decays to zero around the boundaries. Alternatively, the DIC region-of-interest ($\Gamma_{\text{DIC}}$) could be suitably redefined.  Second, when a force-matching term is present, care must be taken to properly balance it with the displacement-matching piece, which can be challenging when the noise in either term is not normally-distributed.

There are also a few technical details related to the nature of the real DIC measurements that we avoided in our presentation. First, in local (subset-based) and global (FE-based) DIC formulations, it is often the case that the basis that represents the DIC displacements is not the same as that used in the FE model, which necessitates interpolation between the DIC point cloud (local) or FE basis (global) to the FE basis for the model. In local DIC the point cloud of measurements is typically very dense, such that interpolation error is unlikely to be a problem unless the FE mesh is extremely fine. Integrated DIC often uses a global formulation where the same mesh is used for both the DIC and model calibration components of the formulation, but projection operators between meshes can be introduced into the inverse formulation if distinct bases are desired. The DIC algorithm can be viewed as low-pass filter on the true deformation field in both a spatial and temporal sense. Consequently, measurements with high spatial gradients and/or rapid temporal variations may be biased, and these limitations should be taken into account when designing aspects of the mechanical test (e.g.\ motion of grips, image acquisition rate). Lastly, the DIC algorithm itself has several user-defined parameters such as subset or element size and basis function order that can affect the amount of random and bias errors present in the DIC measurements.

In this work we treated the BCs on the specimen as known and imposed them directly in the equilibrium equation, but in real experiments choosing what BCs to utilize in the forward model can be difficult. They can be idealized, but this choice can lead to errors when the experimental configuration deviates from the assumed BCs (which is often the case even in well-controlled experiments). The DIC data provides a source of Dirichlet boundary conditions on the DIC surface, and in the case of thin planar specimens, it may be reasonable to extend the in-plane displacements through the thickness in a 3D FE model.  Alternatively, for a thin sample a plane stress model may be appropriate, but its use would require a modified version of the forward formulation that incorporates the plane stress condition as a constraint. For thicker specimens, another option is to set up two back-to-back stereo DIC systems to measure displacements on both sides of a planar sample so that displacement measurements can be interpolated through the thickness of the specimen, thus providing all components of the displacement BCs over the Dirichlet surface \cite{jones2021anisotropic}. Finally, as mentioned in the previous paragraph, while DIC measurements are typically high-quality, they are not perfect and imposing them directly may introduce errors in the forward model.

The final modification is the relaxation of the requirement for the DIC measurements and FE model to share the same temporal discretization. In experiments images are typically acquired at a fixed rate. In many FE codes the nonlinear solver for the forward model is adaptive such that the time step can grow or shrink to improve the robustness of the code and promote computational efficiency. The temporal discretization of the FE model should be aligned with the DIC measurements in time, but it could be allowed to have additional steps. An alternative approach would be to perform temporal interpolation on the DIC data. The derivative of the objective function with respect to the displacement degrees of freedom would be zero for the steps that contain no DIC data.

\section{Conclusion}
\label{sec:conclusion}

In this paper we have described an approach for model calibration from DIC data based on a PDE-constrained optimization formulation with a finite element discretization in which the gradient is obtained through forward or adjoint sensitivity analyses. A challenge in the calibration of elastoplastic constitutive models is the treatment of the coupled nature of the equilibrium PDE, which is satisfied in a global sense, and local equations that govern the time evolution of the constitutive equation. While other studies in the topology optimization literature have considered similar coupled constraints, our reliance on AD to compute the derivatives needed for the forward problem and gradient calculation is novel and frees us from their laborious analytical derivation.

We have shown that both the forward and adjoint sensitivities-based algorithms for computing the gradient are accurate and dramatically reduce the cost of solving the inverse problem relative to using finite differences. Further, we have shown that our computational implementation is suitable for large-scale finite element models on HPC platforms. Promising directions for future work include increasing the computational efficiency of the forward sensitivities approach for use in large-scale problems with many load steps and applying an extended version of the inverse formulation to actual DIC data.

Finally, while computationally-efficient methods for computing gradients of quantities of interest are of course useful for PDE-constrained optimization problems like model calibration or topology optimization, they can also be used to produce more accurate surrogate models than those built using function evaluations alone. Similarly, the adjoint-based sensitivity analysis described in this paper could be modified for use in goal-oriented error estimation. Thus, the methods presented here may find application in areas other than the model calibration field focused on in this work.

\section*{Acknowledgments and Funding}

Supported by the Laboratory Directed Research and Development program at Sandia National Laboratories, a multimission laboratory managed and operated by National Technology and Engineering Solutions of Sandia LLC, a wholly owned subsidiary of Honeywell International Inc. for the U.S. Department of Energy’s National Nuclear Security Administration under contract DE-NA0003525. This paper describes objective technical results and analysis. Any subjective views or opinions that might be expressed in the paper do not necessarily represent the views of the U.S. Department of Energy or the United States Government.

The authors thank Dr. Elizabeth Jones and Dr. Brian Lester for their thorough and thoughtful reviews that led to many improvements in this manuscript.

\section*{Data Availability}

The data that support the findings of this study are available from the corresponding author upon reasonable request.

\appendix

\section{Optimization Details for the Numerical Examples}
\label{sec:cvg_info}

This appendix contains plots of the objective function histories and a table of
function and gradient evaluations for each of the numerical examples presented
in section \ref{sec:results}. The problems in sections 4.1, 4.2, 4.3, are
referred to as circle, large-scale, and cruciform, respectively. We only plot the objective function histories obtained from the adjoint approach, as those obtained with the forward sensitivities approach were
identical and the finite difference FEMU results were extremely similar.

\begin{figure}[ht!]
\centering
\includegraphics[width=0.75\textwidth]{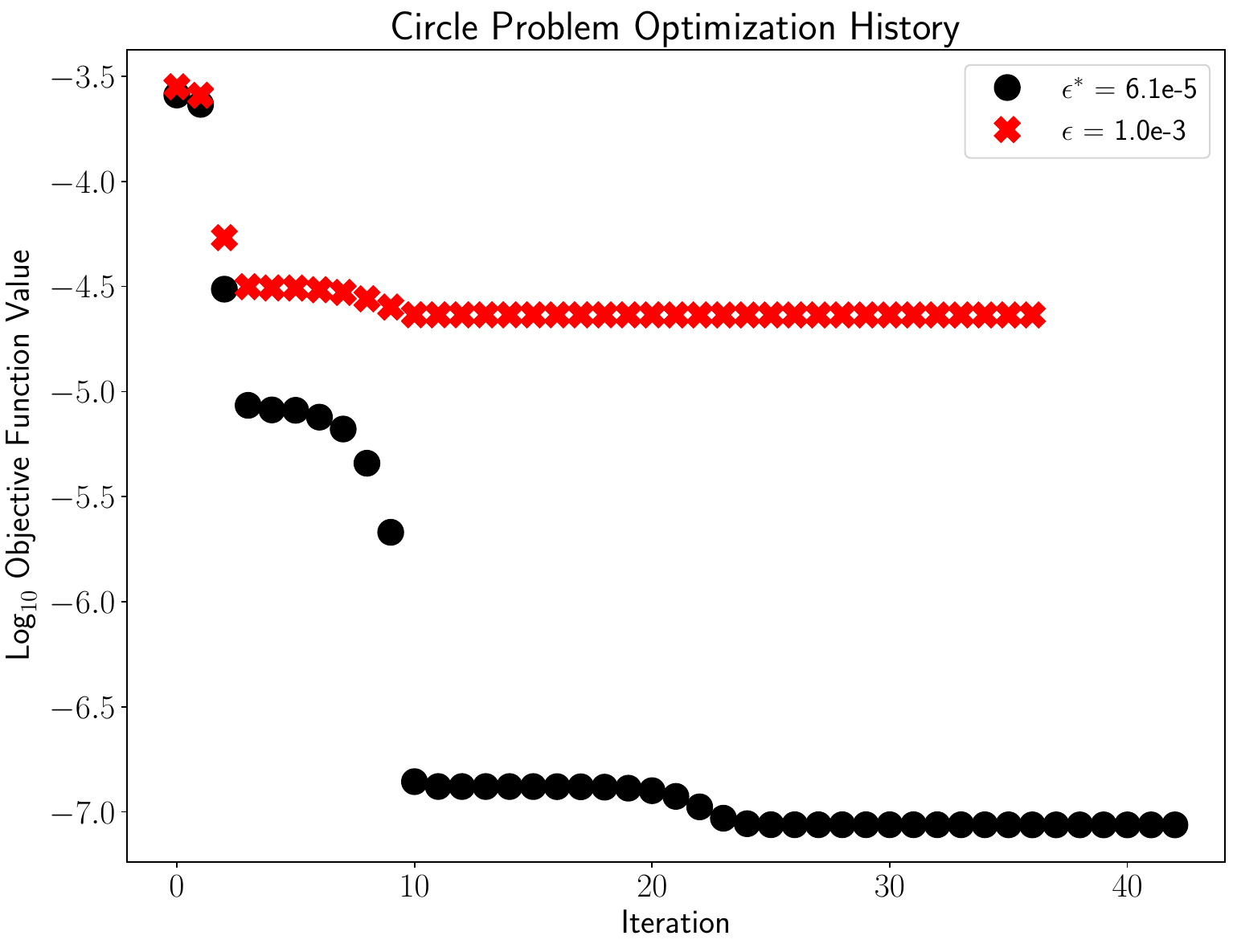}
\caption{Circle example convergence.}
\end{figure}

\begin{figure}[ht!]
\centering
\includegraphics[width=0.75\textwidth]{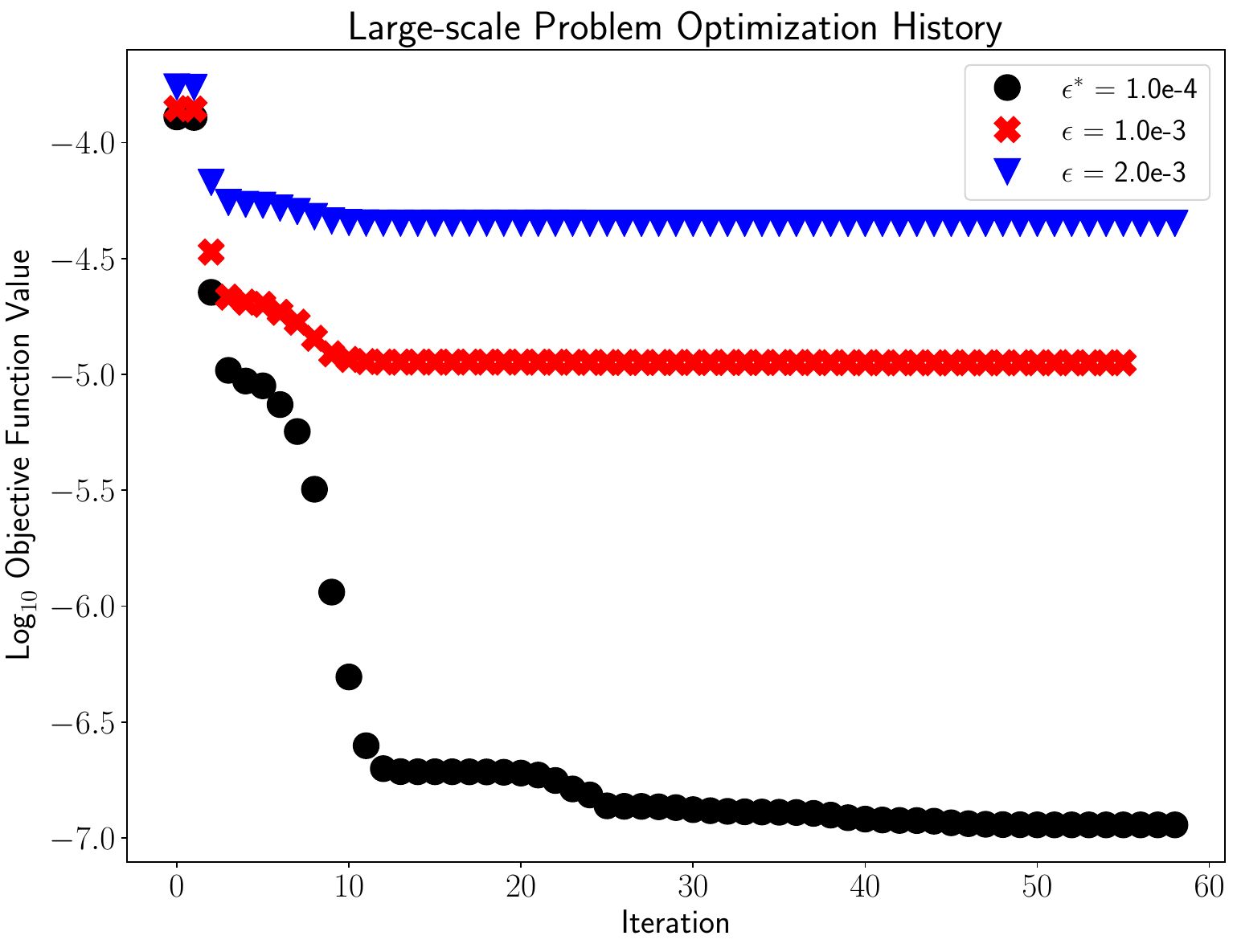}
\caption{Large-scale example convergence.}
\end{figure}

\begin{figure}[ht!]
\centering
\includegraphics[width=0.75\textwidth]{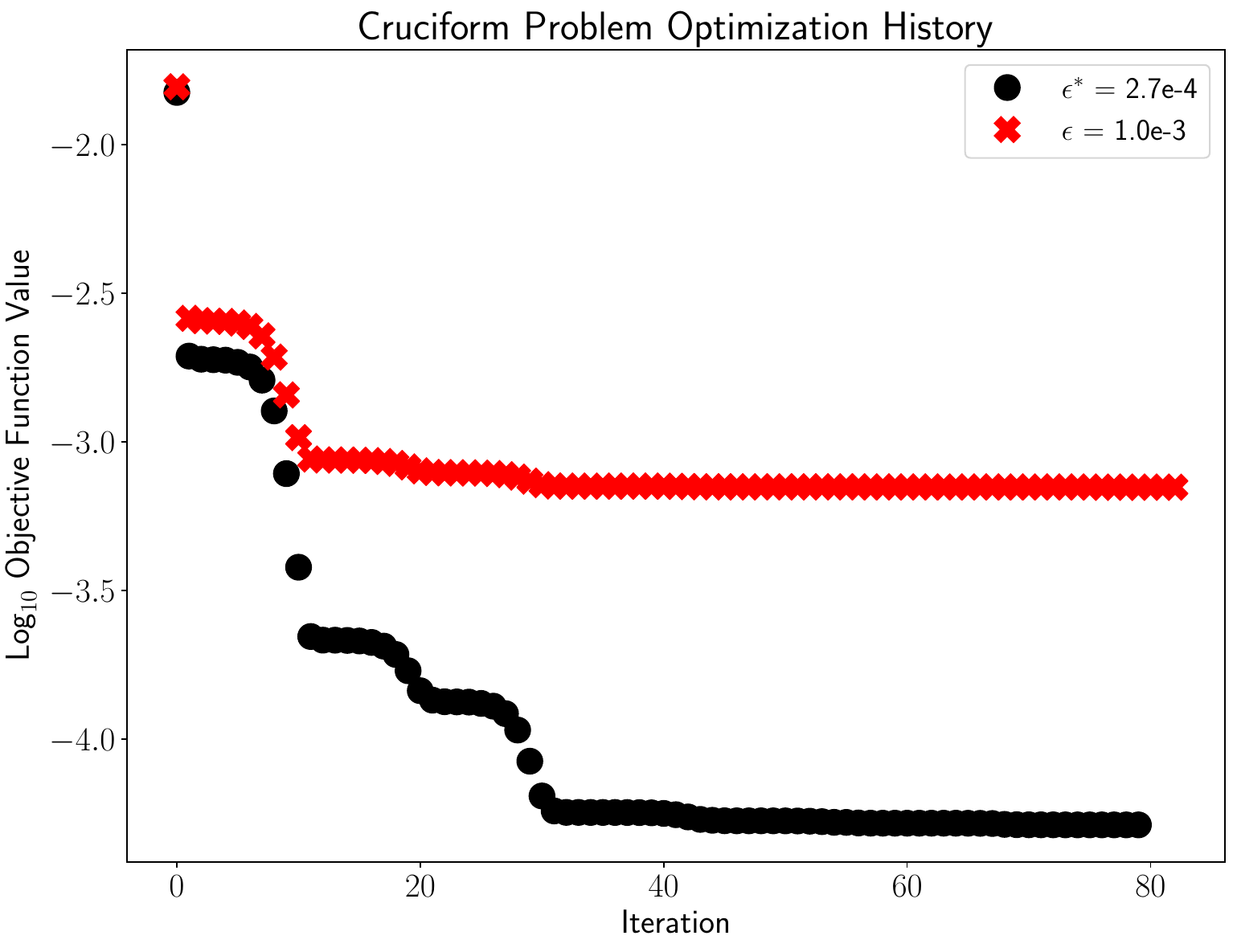}
\caption{Cruciform example convergence.}
\end{figure}

\begin{table}[ht!]
\centering
\begin{tabular}{ | c | c | c | }
\hline
& $\mathcal{J}$ Evaluations & $\frac{d \mathcal{J}}{d \bs{\beta}}$ Evaluations \\ \hline
\multicolumn{3}{|c|}{Circle Problem} \\ \hline
$\epsilon^*_{\text{noise}} =  \num{6.1e-5}$ & 44 & 43 \\ \hline 
$\epsilon_{\text{noise}} =  \num{1.0e-3}$ & 39 & 37 \\ \hline
\multicolumn{3}{|c|}{Large-scale Problem} \\ \hline
$\epsilon^*_{\text{noise}} =  \num{1.0e-4}$ & 64 & 59 \\ \hline
$\epsilon_{\text{noise}} =  \num{1.0e-3}$ & 68 & 56 \\ \hline
$\epsilon_{\text{noise}} =  \num{2.0e-3}$ & 72 & 59 \\ \hline
\multicolumn{3}{|c|}{Cruciform Problem} \\ \hline
$\epsilon^*_{\text{noise}} =  \num{2.7e-4}$ & 88 / 91 & 80 / 77 \\ \hline
$\epsilon_{\text{noise}} =  \num{1.0e-3}$ & 89 / 69 & 83 / 60 \\ \hline
\end{tabular}
\caption{Number of objective function $\mathcal{J}$ and gradient $\frac{d \mathcal{J}}{d \bs{\beta}}$ evaluations for each of the numerical examples. The evaluations reported for the forward sensitivities and adjoint methods (FS-ADJ) were identical but distinct from the FEMU results in the cruciform example, so they are reported separately (i.e.\ FS-ADJ / FEMU). A comparison to FEMU was not performed for the large-scale example.}
\label{table:f_and_g_counts}
\end{table}

\bibliographystyle{plainurl}
\bibliography{coupled_finite_2020.bib}

\end{document}